\documentclass[twocolumn,aps,prd,superscriptaddress,nobalancelastpage,floatfix]{revtex4-1}

\usepackage[T1]{fontenc}
\usepackage[latin9]{inputenc}
\usepackage{lmodern}

\usepackage{amsmath}
\usepackage{amssymb}
\usepackage{amsfonts}
\usepackage{bbold}
\usepackage{graphicx}
\usepackage{bbm}	
\usepackage{tikz}
\usepackage{tikz-cd}

\newtheorem{proposition}{Result}

\usepackage{etoolbox}

\usepackage[dvipsnames]{xcolor}
\usepackage[colorlinks=true,citecolor=blue,linkcolor=red,breaklinks]{hyperref}

\definecolor{URLColor}{HTML}{0B6E4F}   

\makeatletter
\@ifundefined{textcolor}{}{
 \definecolor{BLACK}{gray}{0}
 \definecolor{WHITE}{gray}{1}
 \definecolor{RED}{rgb}{1,0,0}
 \definecolor{GREEN}{rgb}{0,1,0}
 \definecolor{BLUE}{rgb}{0,0,1}
 \definecolor{CYAN}{cmyk}{1,0,0,0}
 \definecolor{MAGENTA}{cmyk}{0,1,0,0}
 \definecolor{YELLOW}{cmyk}{0,0,1,0}
}
\makeatother

\usepackage{tikz-cd}
\usepackage{tikz}
\usetikzlibrary{decorations.pathreplacing,calligraphy,decorations.markings,calc}

\definecolor{tensor}{rgb}{1,1,1}
\definecolor{isometry}{rgb}{1,1,1}
\definecolor{invertibleM}{rgb}{1,1,1}
\definecolor{vector}{rgb}{1,1,1}
\definecolor{unitary}{rgb}{0.8,0.8,1}
\definecolor{gate}{rgb}{1.0,1.0,1.0}

\newcommand{\ATensorLPDO}[2]{
	\begin{scope}[shift={(#1)}]
		\draw (-1.6,0) -- (1.6,0);
		\draw(-0.6,1.2) -- (-0.6,0);
		\draw[red] (0.6,1.2) -- (0.6,0);
		\filldraw[fill=tensor] (-0.9,-0.6) -- (-0.9,0.6) -- (0.9,0.6) -- (0.9,-0.6) -- (-0.9,-0.6);
		\draw (0,0) node {\scriptsize #2};
	\end{scope}
}

\newcommand{\AdagTensorLPDO}[2]{
	\begin{scope}[shift={(#1)}]
		\draw (-1.6,0) -- (1.6,0);
		\draw(-0.6,-1.2) -- (-0.6,0);
		\draw[red] (0.6,-1.2) -- (0.6,0);
		\filldraw[fill=tensor] (-0.9,-0.6) -- (-0.9,0.6) -- (0.9,0.6) -- (0.9,-0.6) -- (-0.9,-0.6);
		\draw (0,0) node {\scriptsize #2};
	\end{scope}
}

\newcommand{\wTensora}[2]{
	\begin{scope}[shift={(#1)}]
		\draw[red] (0,-0.3) --(0,1.3);	   \filldraw[fill=isometry,shift={(0,0)}](0,0.5) circle (0.6);
		\draw (0,0.5) node {\scriptsize #2};
	\end{scope}
}

\newcommand{\ATensorMPS}[2]{
	\begin{scope}[shift={(#1)}]
		\draw (-1.2,0) -- (1.2,0);
		\draw (0,1.2) -- (0,0);
		\filldraw[fill=tensor] (-0.6,-0.6) -- (-0.6,0.6) -- (0.6,0.6) -- (0.6,-0.6) -- (-0.6,-0.6);
		\draw (0,0) node {\scriptsize #2};
	\end{scope}
}

\newcommand{\AdagTensorMPS}[2]{
	\begin{scope}[shift={(#1)}]
		\draw (-1.2,0) -- (1.2,0);
		\draw (0,-1.2) -- (0,0);
		\filldraw[fill=tensor] (-0.6,-0.6) -- (-0.6,0.6) -- (0.6,0.6) -- (0.6,-0.6) -- (-0.6,-0.6);
		\draw (0,0) node {\scriptsize #2};
	\end{scope}
}

\newcommand{\xTensor}[2]{
	\begin{scope}[shift={(#1)}]
		\draw (-1,0) --(1,0);	   
        \filldraw[fill=invertibleM] (0,0) circle[radius=0.6]; 
    \node at (0,0) {\scriptsize #2}; 
	\end{scope}
}

\newcommand{\wTensor}[2]{
	\begin{scope}[shift={(#1)}]
		\draw (0,-0.3) --(0,1.3);	   \filldraw[fill=isometry,shift={(0,0)}](0,0.5) circle (0.58);
		\draw (0,0.5) node {\scriptsize #2};
	\end{scope}
}

\makeatother

\usepackage{babel}

\definecolor{ZM}{rgb}{.5,0,.5}
\definecolor{FP}{rgb}{1,0,0}
\definecolor{SD}{rgb}{0,1,0}

\begin{document}

\title{Mixed-state topological transitions with hidden long-range correlations}

\author{Ze-Min Huang}
\affiliation{Institute for Theoretical Physics, University of Cologne, 50937 Cologne, Germany}

\author{Frank Pollmann}
\affiliation{Technical University of Munich, TUM School of Natural Sciences, Physics Department, James-Franck-Stra{\ss}e 1,
85748 Garching, Germany}
\affiliation{Munich Center for Quantum Science and Technology (MCQST), Schellingstr. 4, 80799 M{\"u}nchen, Germany}

\author{Sebastian Diehl}
\affiliation{Institute for Theoretical Physics, University of Cologne, 50937 Cologne, Germany}

\begin{abstract}
We study a decoherence-driven transition from a one-dimensional symmetry-protected topological state to a trivial product state. The decoherence acts onsite while preserving the protecting symmetry on average. Although conventional local correlations remain short-ranged across the transition, we uncover hidden nondecaying long-range correlations in a symmetry-twisted ensemble that weights states according to their symmetry charge. At the transition, the twisted correlation length diverges and the twisted partition function becomes singular. These signatures can be directly extracted from symmetry-resolved measurement outcomes, making the hidden critical behavior accessible on current quantum platforms at mesoscopic system sizes. We demonstrate this through numerical simulations of the measurement process.
\end{abstract}
\maketitle
\section{Introduction}
Landau-Ginzburg theory is the traditional framework for classifying phases of matter, according to patterns of symmetry breaking and their associated local order parameters. Topological phases go beyond this by identifying distinct phases even when they share the same unbroken symmetries~\cite{wen2004oxford, fradkin2013cambridge}. A prominent example is the symmetry-protected topological (SPT) phase~\cite{hassan2010rmp,pollmann2010prb,qi2011rmp, pollmann2012prb, schuch2011prb,chen2011science, chen2011prb, senthil2015arcmp, pollmann2017oxford}, which cannot be continuously deformed into a trivial state without breaking the protecting symmetry or crossing a phase transition at which the correlation length diverges. The SPT order is not encoded in any local observable, but instead manifests through nonlocal quantities such as selection rules of string order~\cite{nijs1989prb,kennedy1992prb,garcia2008prl,pollmann2012prb} and quantized responses to flux insertion~\cite{kitaev2001pu,hatsugai2006jpsj,hirano2008prb,zaletel2014njp, shiozaki2017jhep,vanderbilt2018cambridge}. This nonlocality captures the global properties of the quantum state and thus reflects the topological character. Yet this same feature also poses a major experimental challenge, as conventional local probes offer only limited access to such global information, making SPT phases difficult to detect in traditional condensed-matter platforms.

Rapid advances in quantum technologies have begun to overcome this limitation by providing highly controllable analog and digital quantum simulators~\cite{nori2014rmp,altmann2021prxq,gammonsmoth2026np}. These platforms enable the preparation of diverse topological states \cite{choo2018prl,smith2022prr,smith2023prxq,satzinger2021science,semeghini2021science,sompet2022nature,iqbal2024nature}, while providing microscopic access to observables that are difficult to probe in conventional materials, such as nonlocal string-order parameters~\cite{semeghini2021science,sompet2022nature} and site-resolved charge distributions~\cite{gross2017science}. However, no realistic quantum simulator is perfectly isolated from its environment. Coupling to the environment introduces decoherence and can exchange symmetry charge with the surroundings, converting a prepared pure state into a mixed state and potentially obscuring or destroying its topological character. The interplay between decoherence and topology has therefore attracted considerable attention~\cite{degroot2022quantum,huang2022prb,ma2023prx,zang2024prl,huang2025prl,huang2025prl_mspt,sohal2025prxq,ellison2025prxq,lee2025quantum,mao2025cpl,guo2025prx}, with a key organizing principle being the distinction between strong and weak symmetries~\cite{buca2012njp}. Strong symmetry precludes system-environment symmetry-charge exchange, whereas weak symmetry preserves symmetry only at the ensemble level. This distinction underlies the notion of average SPT phases~\cite{ma2023prx}, which require at least one strong symmetry, and intrinsically mixed-state phenomena such as strong-to-weak spontaneous symmetry breaking (SWSSB)~\cite{sala2024prb,lessa2025prxq}.

Meanwhile, given the broad relevance of symmetry-charge exchange between a system and its environment, a distinct yet fundamental question arises: What is the fate of topological responses under weak symmetry alone? Both string-order selection rules and quantized responses have recently been shown to remain robust in this setting~\cite{huang2025prl_mspt}. Yet, under sufficiently strong noise, this response eventually changes. In pure states, such a change along a continuous, symmetry-preserving path necessarily requires a diverging correlation length~\cite{pollmann2010prb,chen2013prb,cirac2021rmp}. Under weakly symmetric decoherence, however, whether this remains true is far less clear, especially in one dimension, where conventional finite-temperature criticality is absent, making a divergent conventional correlation length unexpected. What, then, becomes critical, and how can it be observed?

In this work, we address this question for decoherence-driven transitions
from SPT states to trivial states and demonstrate unconventional critical
behavior in a symmetry-twisted ensemble. The associated twisted partition
function $Z_g$ serves as our central object. 
When the topological response changes, $Z_g$ becomes nonanalytic in the thermodynamic limit, with the twisted free-energy density $-\ln|Z_g|/N$ ($N$ being the number of sites) developing a cusp. At the same transition, the correlation length of the symmetry-twisted ensemble diverges, and the corresponding connected correlations become nondecaying, rather than algebraically decaying, at large distances. By contrast, the untwisted correlation length remains finite, and conventional local correlations stay short-ranged. Thus, the long-range correlations exist exclusively in the symmetry-twisted sector. Also, this combination of a singular twisted free energy and a divergent twisted correlation length is reminiscent of the matrix-product-state (MPS) quantum
phase transitions~\cite{wolf2006prl,jones2021prr}. In this respect, our transition
realizes a decoherence-driven mixed-state analogue of such unconventional
critical behavior. Although invisible to conventional local probes, these symmetry-twisted signatures are accessible on current quantum platforms through repeated projective measurements on independently prepared copies of the state, as we demonstrate numerically in concrete models.

This paper is organized as follows. In Sec.~\ref{sec:review}, we review topological responses in pure and mixed SPT states. In Sec.~\ref{sec:result}, we show that mixed-state topological transitions are accompanied by a divergent correlation length and cusp-like singularities in symmetry-twisted ensembles. In Sec.~\ref{sec:ex1_cluster}, we illustrate these results using the cluster state and numerically investigate their observability through quantum-microscope measurements. In Sec.~\ref{sec:ex2_dHaldane}, we further test these predictions numerically in a spin-1 chain.

\section{Review of Topological Responses in Symmetry-Protected Topological Pure and Mixed States \label{sec:review}}
We begin by reviewing order parameters for symmetry-protected topological (SPT) phases in pure and mixed states, in a form suited to our later constructions. These include quantized responses and selection rules for string-order parameters. 
Since every one-dimensional SPT phase admits an MPS representative, we formulate these diagnostics directly in the MPS framework~\cite{cirac2021rmp}.

\subsection{Pure topological state protected by symmetry}
Consider a pure bosonic SPT state $|\psi\rangle$ of a one-dimensional $N$-site spin chain protected by an on-site Abelian finite group $G$, i.e., $U_g|\psi\rangle\sim|\psi\rangle$ up to a phase factor, with $g\in G$ and $U_g=u_g^{\otimes N}$. For these area-law entangled states, the wavefunction can be efficiently represented using MPS,
\begin{eqnarray}
|\psi(X)\rangle &=&\sum_{\{i_j\}}\text{tr}(X\ A^{i_1}A^{i_2}...A^{i_N})|i_1,i_2,...,i_N\rangle\nonumber\\
&=&	\begin{tikzpicture}[scale=.45, baseline={([yshift=0.5ex]current bounding box.center)}, thick] 
      
		\ATensorMPS{0, 1}{$A$}
        \ATensorMPS{2, 1}{$A$}
        \draw[shift={(0,0)},dotted] (2.8,1) -- (5,1);
        \ATensorMPS{6, 1}{$A$}
       
         \draw[shift={(0,0)}] (7.2,0.6) arc(-90:90:0.2);
		\draw [decorate,
    	decoration = {calligraphic brace,mirror}] (-0.5,-0) --  (6.8,-0);
		\draw (2.7,-0.5) node {\scriptsize $N$};
         \xTensor{-1.55, 1}{$X$}
          \draw[shift={(0,2.8)}] (-2.5,-1.8) arc(90:270:0.2);
           
	\end{tikzpicture}
\end{eqnarray}
where $|i_j\rangle$ spans the physical Hilbert space at site $j$, $A$ is a rank-3 tensor with $A^{i_j}$ a matrix in the virtual space, and "$\text{tr}$" ("$\text{Tr}$") denotes tracing over the virtual (physical) indices. The matrix $X$ encodes boundary conditions, and $|\psi(\mathbb{1})\rangle$ is denoted as $|\psi\rangle$ unless stated otherwise.

These states are categorized into distinct equivalence classes according to the symmetry action on the virtual space, and states belonging to different classes cannot be adiabatically connected without breaking the symmetry~\cite{pollmann2010prb,chen2011prb,schuch2011prb}. Specifically, the symmetry transformation of the tensor $A$, which, up to a gauge choice, takes the form~\cite{perez_garcia2007,cirac2021rmp} (see Appendix~\ref{app:fundamental_thm} for details)
\begin{equation}
\begin{tikzpicture}[scale=.45, baseline={([yshift=-1.3ex]current bounding box.center)}, thick]
    \ATensorMPS{0,0}{$A$}
    \wTensor{0,0.9}{$u_g$}  
  \end{tikzpicture}=e^{i\theta_g}
\begin{tikzpicture}[scale=.45, baseline={([yshift=-1.3ex]current bounding box.center)}, thick]
    \ATensorMPS{0,0}{$A$}
     \xTensor{1.6, 0}{$V_g^\dagger$}
     \xTensor{-1.6, 0}{$V_g$}
  \end{tikzpicture},\label{eq:injectivity_pure}
\end{equation}
which follows from the short-range entangled nature of SPT states~\cite{schuch2011prb,cirac2021rmp}. The matrices $V_g$ form a projective representation of the symmetry group, i.e., $V_{g_1}V_{g_2}=V_{g_1g_2}\omega(g_1,g_2)$, where $g_1, g_2\in G$, and $\omega$ is a phase factor~\cite{chen2011prb, schuch2011prb, chen2013prb}. In particular, the representation is defined only up to a gauge transformation, $V_g\rightarrow V_ge^{i\phi(g)}$ and $\omega(g_1,g_2)\rightarrow \omega(g_1,g_2)e^{-i\phi(g_1 g_2)+i\phi(g_1)+i\phi(g_2)}$, so those related by such a transformation are gauge equivalent and therefore belong to the same cohomology class~\cite{chen2011prb, schuch2011prb, chen2013prb}. 
Additionally, Eq.~\eqref{eq:injectivity_pure} directly implies that a $g_1$-flux insertion is implemented by choosing $X=V_{g_1}$~\cite{zaletel2014njp,zaletel2013prl}. Indeed, this identity allows the symmetry twist acquired by particles winding around the spatial circle, in analogy to the Aharonov-Bohm effect, to be transferred to the virtual bond; see Appendix~\ref{app_sec:flux_insertion} for details.

Physically, distinct SPT classes can be distinguished by two observables: (i) The quantized responses to flux insertion,
  \begin{equation}
e^{i\mathcal{Q}(g_1, g_2)}\equiv \langle \psi(V_{g_1})|U_{g_2}|\psi(V_{g_1})\rangle/\langle \psi|U_{g_2}|\psi\rangle,
\label{eq:pure_quantized}
\end{equation}
which measures the charge response of the flux-inserted state $|\psi(V_{g_1})\rangle$ relative to the original state $|\psi\rangle$.
(ii) The string order parameter, i.e., 
 \begin{equation}
\mathcal{S}^{(n)}(g_2, \chi_{g_1}^L, \chi_{g_1}^R)\equiv \langle \psi| \hat{\mathcal{S}}(g_2, \chi_{g_1}^L, \chi_{g_1}^R)|\psi\rangle/|\langle \psi|U_{g_2}|\psi\rangle|^{l/N}.
\end{equation}
Here, the operator $\hat{\mathcal{S}}=\chi_{g_1}^L\otimes (u_{g_2})^{\otimes l}\otimes \chi_{g_1}^R$ creates a symmetry-twisted domain of length $l$, terminated by the endpoint operators $\chi_{g_1}^L$ and $\chi_{g_1}^R$. The denominator $|\langle\psi|U_{g_2}|\psi\rangle|^{l/N}$ normalizes the string order parameter by removing the extensive contribution of the symmetry twist over the fraction $l/N$ of the system, and equals unity for a normalized symmetric pure state. We may therefore omit it for simplicity and denote the resulting unnormalized string order parameter by $\mathcal{S}$.
In particular, as shown below, the charge response $e^{i\mathcal{Q}(g_1,g_2)}$ coincides with the topological invariant~\cite{zaletel2014njp} that distinguishes different cohomology classes and can therefore change only when the relevant gap closes.
Meanwhile, the string order parameter detects this invariant through its selection rule~\cite{pollmann2012prb},
\begin{equation}\label{eq:selection_normalized_sop}
\mathcal{S}^{(n)}(g_{2},\chi_{g_{1}}^{L},\chi_{g_{1}}^{R})|_{l\rightarrow\infty}=\begin{cases}
\mathcal{O}\left(1\right),\ \text{if}\ e^{i\mathcal{Q}(g_{1},g_{2})}= e^{i\phi(g_{1})}\\
0,\quad\text{if}\ e^{i\mathcal{Q}(g_{1},g_{2})}\neq e^{i\phi(g_{1})}
\end{cases}.
\end{equation}
Here, the $\mathcal{O}(1)$ value holds generically in the absence of fine tuning, while
$e^{\pm i \phi(g_1)}$ is the symmetry charge of the end-point operator $\chi^{L/R}$, i.e.,  $u_{g_1}\chi_{g_1}^{L/R} u_{g_1}^\dagger = e^{\pm i \phi(g_1)}\chi_{g_1}^{L/R}$.

We now establish the  observable consequences of the above. To this end, we define the symmetry-twisted transfer matrix $T(g_2)$, i.e.,
\begin{equation}
T(g_2)\equiv \begin{tikzpicture}[scale=.45, baseline={([yshift=0.5ex]current bounding box.center)}, thick]
    \AdagTensorMPS{0,3}{$A^*$}
    \ATensorMPS{0,0}{$A$}
    \wTensor{0,1}{$u_{\scriptscriptstyle g_2}$}
\end{tikzpicture}
\end{equation}
whose contraction yields the $g_2$-twisted partition function
\begin{equation}\label{eq:symmetry_Zg}
Z_{g_2}\equiv \text{Tr}\left(\rho U_{g_2}\right)=\text{tr}\left[T^N(g_2)\right],
\end{equation}
with $\rho$ the density matrix, thereby defining the symmetry-twisted ensemble used below.
Crucially, the transfer matrix $T(g_2)$ commutes with the flux operator, i.e., 
\begin{equation}\label{eq:pure_commute}
[T(g_2), 
\centering
\begin{tikzpicture}[scale=.45, baseline={([yshift=0.5ex]current bounding box.center)}, thick]
   \xTensor{0, 0}{$V_{g_1}$}
    \xTensor{0, 2}{$V^*_{g_1}$}
\end{tikzpicture}\ ]=0,
\end{equation}
as follows directly from the push-through identity in Eq.~\eqref{eq:injectivity_pure} for Abelian $G$, i.e., 

\begin{equation}
[u_{g_1}, u_{g_2}]=0\implies
\begin{tikzpicture}[scale=.45, baseline={([yshift=0.5ex]current bounding box.center)}, thick]

    \AdagTensorMPS{0,3}{$A^*$}
    \ATensorMPS{0,0}{$A$}
    \wTensor{0,1}{$u_{\scriptscriptstyle g_2}$}
    \xTensor{-1.6, 3}{$V_{g_1}^*$}
    \xTensor{-1.6, 0}{$V_{g_1}$}

    \xTensor{1.6, 3}{$V_{g_1}^T$}
    \xTensor{1.6, 0}{$V_{g_1}^\dagger$}
\end{tikzpicture}=
\begin{tikzpicture}[scale=.45, baseline={([yshift=0.5ex]current bounding box.center)}, thick]
    \AdagTensorMPS{0,3}{$A^*$}
    \ATensorMPS{0,0}{$A$}
    \wTensor{0,1}{$u_{\scriptscriptstyle g_2}$}
\end{tikzpicture}.
\end{equation}
This commutation relation Eq.~\eqref{eq:pure_commute}, together with the spectral gap of $T(g_2)$, underlies the robustness of the topological response: neither the quantized response nor the associated selection rule can change as long as the gap remains open. 

Specifically, the spectral decomposition of $T(g_2)$ is
\begin{equation}
\begin{split}
T(g_2)
=\sum_{n}\lambda^{(n)}_{g_2}
\begin{array}{c}
\begin{tikzpicture}[scale=.5, baseline={([yshift=0ex]current bounding box.center)}, thick]
\draw[rounded corners] (0,2.8)--(0.55,2.8)--(0.55,0)--(0,0);
\filldraw[fill=vector] (0.05,1.0) rectangle (1.05,1.8);
\draw (0.55,1.4) node {\scriptsize $R_{g_2}^{n}$};
\draw[rounded corners] (2.4,2.8)--(1.85,2.8)--(1.85,0)--(2.4,0);
\filldraw[fill=vector] (1.35,1.0) rectangle (2.35,1.8);
\draw (1.85,1.4) node {\scriptsize $L_{g_2}^{n}$};
\end{tikzpicture}
\end{array}
=\sum_{n}\lambda_{g_2}^{(n)}|R_{g_2}^{n})(L_{g_2}^{n}|
\end{split},
\end{equation}
where $|R_{g_2}^n)$ and $(L_{g_2}^n|$ denote the right and left normalized eigenvectors of $T(g_2)$, respectively. Here and below, for presentation simplicity, we assume that $T(g_2)$ is diagonalizable;
the non-diagonalizable case follows analogously from the Jordan decomposition,
without changing the exponential scales relevant below. The eigenvalues $\lambda_{g_2}^{(n)}\in \mathbb{C}$ are ordered by decreasing magnitude $|\lambda_{g_2}^{(n)}|$ as $n=0,1,...$; in degenerate cases, a basis is chosen implicitly within each eigenspace. The spectral gap is then defined as
\begin{align}\label{eq:gap}
    \Delta({g_2})\equiv |\lambda_{g_2}^{(0)}|-|\lambda_{g_2}^{(1)}|,
\end{align} 
and crucially controls the correlation length of the $g_2$-twisted ensemble. In particular, the correlation length is defined through the asymptotic decay of connected correlation functions of local operators $O_i$, i.e., 
\begin{equation}
\left|\langle O_i O_j\rangle_{g_2}
-\langle O_i\rangle_{g_2}\langle O_j\rangle_{g_2}\right|_{|i-j|\gg 1}
\sim e^{-|i-j|/\xi(g_2)},
\end{equation}
where the expectation value in the $g_2$-twisted ensemble is
\begin{equation}\label{eq:expectation_twisted_ensemble}
\langle \cdots \rangle_{g_2}
\equiv
\frac{\operatorname{Tr}\left[\rho U_{g_2}(\cdots)\right]}
{\operatorname{Tr}\left(\rho U_{g_2}\right)}.
\end{equation}
In terms of the transfer-matrix spectrum, this correlation length is 
\begin{equation}\label{eq:correlation_length}
\xi(g_2)= -\frac{1}{\ln |\lambda_{g_2}^{(1)}/\lambda_{g_2}^{(0)}|}=
-\frac{1}{\ln[1-\Delta(g_2)/ |\lambda_{g_2}^{(0)}|]},
\end{equation}
and therefore diverges when the symmetry gap closes,  
\begin{equation}
|\Delta(g_2)/\lambda_{g_2}^{(0)}|\ll1:\ \xi(g_2)\propto \frac{1}{\Delta(g_2)}.
\end{equation}
Equation~\eqref{eq:correlation_length} can then be derived directly from the tensor-network representation. By considering the $|i-j|\gg 1$ limit and assuming a unique subleading eigenvalue $|\lambda_{g_2}^{(1)}|$ for simplicity, the twisted two-point correlation function reduces to,
\begin{equation}
|\langle O_i O_j\rangle_{g_2}-\langle O_i\rangle_{g_2}\langle O_j\rangle_{g_2}|=|\lambda_{g_2}^{(1)}/\lambda_{g_2}^{(0)}|^{|i-j|}\times\mathcal{O}(1),
\end{equation}
with the $\mathcal{O}(1)$ coefficient given by, 
\begin{eqnarray} \left|
\begin{array}{c}
\begin{tikzpicture}[
    scale=.45,
    baseline={([yshift=0ex]current bounding box.center)},
    thick
]

\ATensorMPS{-0.2,0}{$A$}
\AdagTensorMPS{-0.2,4.2}{$A^*$}

\wTensor{-0.2,0.9}{$u_{g_2}$}
\wTensor{-0.2,2.3}{$O_i$}

\draw[rounded corners]
    (-1.2,4.2) -- (-1.75,4.2)
    -- (-1.75,0) -- (-1.2,0);

\filldraw[fill=vector]
    (-2.25,1.6) rectangle (-1.25,2.6);

\node at (-1.75,2.1)
    {\scriptsize $L_{g_2}^{0}$};

\draw[rounded corners]
    (1.0,4.2) -- (1.55,4.2)
    -- (1.55,0) -- (1.0,0);

\filldraw[fill=vector]
    (1.05,1.6) rectangle (2.05,2.6);

\node at (1.55,2.1)
    {\scriptsize $R_{g_2}^{1}$};

\end{tikzpicture}\times
\end{array}
\begin{array}{c}
\begin{tikzpicture}[
    scale=.45,
    baseline={([yshift=0ex]current bounding box.center)},
    thick
]

\ATensorMPS{-0.2,0}{$A$}
\AdagTensorMPS{-0.2,4.2}{$A^*$}

\wTensor{-0.2,0.9}{$u_{g_2}$}
\wTensor{-0.2,2.3}{$O_j$}

\draw[rounded corners]
    (-1.2,4.2) -- (-1.75,4.2)
    -- (-1.75,0) -- (-1.2,0);

\filldraw[fill=vector]
    (-2.25,1.6) rectangle (-1.25,2.6);

\node at (-1.75,2.1)
    {\scriptsize $L_{g_2}^{1}$};

\draw[rounded corners]
    (1.0,4.2) -- (1.55,4.2)
    -- (1.55,0) -- (1.0,0);

\filldraw[fill=vector]
    (1.05,1.6) rectangle (2.05,2.6);

\node at (1.55,2.1)
    {\scriptsize $R_{g_2}^{0}$};

\end{tikzpicture}
\end{array}\right|.
\label{eq:Oi-leading-contraction}
\end{eqnarray}
For pure states, $\xi(g_2)$ is independent of $g_2$, yielding a single characteristic correlation length shared by all symmetry-twisted ensembles. This follows from the push-through identity in Eq.~\eqref{eq:injectivity_pure}, which makes $T(g_2)$ isospectral for all $g_2$ and hence $|\Delta(g_2)|$ independent of $g_2$. For mixed states, this isospectrality generally fails, allowing $\Delta(g_2)$ to depend on $g_2$; we therefore refer to $\Delta(g_2)$ as the $g_2$-symmetry gap, see below for details.

For short-range-correlated pure SPTs, the gapped spectrum of $T(g_2)$ greatly simplifies the quantized response. In the $N\gg 1$ limit, both $\langle\psi|U_{g_2}|\psi\rangle$ and $\langle\psi(V_{g_1})|U_{g_2}|\psi(V_{g_1})\rangle$ are dominated by the leading eigenvectors of the corresponding transfer matrices, with subleading contributions exponentially suppressed as $\left|\lambda_{g_2}^{(n)}/\lambda_{g_2}^{(0)}\right|^N$, i.e.,
\begin{equation}\label{eq:topological_response_pure}
e^{i\mathcal{Q}(g_1, g_2)}=(L_{g_2}^0|V_{g_1}^*\otimes V_{g_1}|R_{g_2}^0).
\end{equation}
Additionally, the finite gap $\Delta(g_2)$, and hence the uniqueness of the leading eigenstate, ensures that $e^{i\mathcal{Q}(g_1,g_2)}$ furnishes a one-dimensional representation of the cyclic subgroup generated by $g_1$. It therefore takes values in a discrete set and labels distinct equivalence classes, and thus yields a quantized response. 
In turn, any change in the topological response necessitates a closing of the gap $\Delta(g_2)$, thereby driving the correlation length of the $g_2$-twisted ensemble to diverge. For pure states, a change in the topological response along a continuously varying, symmetry-preserving path requires a divergent correlation length already in the untwisted ensemble~\cite{schuch2011prb,cirac2021rmp}, since there exists only one characteristic correlation length $\xi(\mathbb{1})$.

For the string order parameter, a similar reduction to the leading eigenstate of $T(g_2)$ applies when $l\gg 1$, allowing us to derive the selection rule Eq.~\eqref{eq:selection_normalized_sop}. Pictorially, the string order parameter is 
\begin{equation}
\begin{aligned}
\mathcal{S}^{(n)}
&=
\frac{1}{
\left|\langle\psi|U_{g_2}|\psi\rangle\right|^{l/N}
}
\\[-0.5ex]
&\quad\times
\vcenter{\hbox{%
\begin{tikzpicture}[
    scale=.45,
    baseline={([yshift=0.5ex]current bounding box.center)},
    thick
]

    \ATensorMPS{0,0}{$A$}
    \ATensorMPS{2,0}{$A$}
    \ATensorMPS{6,0}{$A$}
    \ATensorMPS{8,0}{$A$}
    \ATensorMPS{12,0}{$A$}

    \draw (-1.2,0) arc (90:270:0.2);
    \draw (13.2,-0.4) arc (-90:90:0.2);

    \AdagTensorMPS{0,3}{$A^*$}
    \AdagTensorMPS{2,3}{$A^*$}
    \AdagTensorMPS{6,3}{$A^*$}
    \AdagTensorMPS{8,3}{$A^*$}
    \AdagTensorMPS{12,3}{$A^*$}

    \draw (-1.2,3) arc (90:270:0.2);
    \draw (13.2,2.6) arc (-90:90:0.2);

    \draw[dotted] (2.8,0) -- (5.0,0);
    \draw[dotted] (2.8,3) -- (5.0,3);
    \draw[dotted] (8.8,0) -- (11.0,0);
    \draw[dotted] (8.8,3) -- (11.0,3);

    \draw (0,0.9) -- (0,2.1);
    \draw (12,1.2) -- (12,1.8);

    \wTensor{0,1}{$\chi_L$}
    \wTensor{2,1}{$u_{g_2}$}
    \wTensor{6,1}{$u_{g_2}$}
    \wTensor{8,1}{$\chi_R$}

    \draw[
        decorate,
        decoration={calligraphic brace,mirror}
    ] (1.35,-0.85) -- (6.65,-0.85);

    \node at (4,-1.35) {\scriptsize $l$};

\end{tikzpicture}%
}} .
\end{aligned}
\end{equation}
Similar to the quantized response, in the limit $l\gg 1$, the transfer-matrix products are dominated by their leading eigenvectors, yielding
\begin{equation}\label{eq:string_leading}
\mathcal{S}^{(n)}
=
\bigl(L_{\mathbb{1}}^{0}\big|T(\chi_L)\big|R_{g_2}^{0}\bigr)
\bigl(L_{g_2}^{0}\big|T(\chi_R)\big|R_{\mathbb{1}}^{0}\bigr).
\end{equation}
where $T(\chi_{L/R})$ denotes the transfer matrix obtained from $T(g_2)$ by replacing $u_{g_2}$ with $\chi_{L/R}$. The selection rule in Eq.~\eqref{eq:selection_normalized_sop} then follows from the fact that both $(L^0_{\mathbb{1}}|T(\chi_L)$ and $|R^0_{g_2})$ are eigenvectors of the flux operator $V_{g_1}^*\otimes V_{g_1}$, i.e., 
\begin{equation}
 (L^0_{\mathbb{1}}|T(\chi_L)\ V_{g_1}^*\otimes V_{g_1} =
e^{i\phi(g_1)}\ 
(L^0_{\mathbb{1}}|T(\chi_L),
\end{equation}
and 
\begin{equation}
V_{g_1}^*\otimes V_{g_1}\ |R^0_{g_2})=
e^{i\mathcal{Q}(g_1,g_2)}\ |R^0_{g_2}),
\end{equation}
so their overlap Eq.~\eqref{eq:string_leading} can be nonzero only if $e^{i\mathcal{Q} (g_1,g_2)}=e^{i\phi(g_1)}$.

\subsection{Mixed topological state protected by symmetry}
We now turn to mixed-state SPTs and demonstrate that their nonlocal order parameters remain robust as long as the symmetry gap remains open. However, unlike in the pure-state case, the string-order selection rule manifests only in the normalized string-order parameter, while remaining obscured in its unnormalized counterpart.

To this end, we proceed as in the pure-state case and work in the MPS representation. Concretely, we focus on mixed states that admit purifications with short-range-correlated MPS representations~\cite{verstrate2004prl, zwolak2004prl, cuevas2013njp},
\begin{equation}
|\Psi\rangle=
	\begin{tikzpicture}[scale=.45, baseline={([yshift=0.3ex]current bounding box.center)}, thick]
        
		\ATensorLPDO{0.3, 0}{$A$}
        
        \ATensorLPDO{5.7, 0}{$A$}

        \draw[shift={(-1.4,0.8)}] (-1.2,-0.8) arc(90:270:0.2);
         \draw[shift={(-1.3,0)}] (8.6,-0.4) arc(-90:90:0.2);
		\draw [decorate,
    	decoration = {calligraphic brace,mirror}] (-0.5,-0.8) --  (6.8,-0.8);
		\draw (2.7,-1.5) node {\scriptsize $N$};  
        \draw[shift={(0,0)},dotted] (2,0) -- (4,0);
        \xTensor{-1.6, 0}{$X$}
	\end{tikzpicture}\ ,
\end{equation}
which is known as the locally-purified-density-operator (LPDO) ansatz~\cite{verstrate2004prl}. Here, the tensor $\begin{array}{c} 
    \begin{tikzpicture}[scale=.4, baseline={([yshift=-5ex]current bounding box.center)}, thick]
    	\ATensorLPDO{0,0}{$A$}
    	\draw (-0.4,1.2) node {\scriptsize $i$};
	\draw (0.9,1.2) node {\scriptsize $a$};
    	\draw (1.8,0) node {\scriptsize $\beta$};
    	\draw (-1.8,0) node {\scriptsize $\alpha$};
    \end{tikzpicture}
\end{array}$ carries both the physical index $i$ (black vertical leg) and the ancilla index $a$ (red vertical leg). As in the pure-state case, the boundary matrix $X$ specifies the boundary condition, including $g$-twisted boundary conditions~\cite{huang2025prl_mspt} (see Appendix~\ref{app_sec:flux_insertion} for details). The system density matrix is then obtained by tracing out the ancillas, $\rho(X)=\mathrm{Tr}_a |\Psi(X)\rangle\langle\Psi(X)|$, with $\rho(\mathbb{1})$ denoted as $\rho$ unless stated otherwise.

Different from pure states, symmetries of mixed states come in two forms~\cite{buca2012njp}: a weak symmetry satisfies $[U_g,\rho]=0$, whereas a strong symmetry satisfies $U_g\rho \sim \rho$, up to a phase factor. For symmetries that admit a local realization in the enlarged Hilbert space including the ancillas, the operator $U_g\otimes U_g^a=(u_g\otimes u_g^a)^{\otimes N}$~\cite{degroot2022quantum} acts on the purified wavefunction as,
\begin{equation}\label{eq:injectivity_weak}
\begin{array}{c}
\begin{tikzpicture}[scale=.45,thick,baseline={([yshift=-9ex]current bounding box.center)}]
			\ATensorLPDO{0,0}{$A$}
			\wTensor{-0.6,0.9}{$u_g$}
            \wTensora{0.6,0.9}{$u_g^a$}
\end{tikzpicture}
\end{array}
=e^{i\theta_g}
\begin{array}{c}
\begin{tikzpicture}[scale=.44,thick,baseline={([yshift=-4ex]current bounding box.center)}]
			\ATensorLPDO{0,0}{$A$}
			\xTensor{-2.2,0}{$V_g$}
            \xTensor{2.2,0}{$V_g^\dagger$}
\end{tikzpicture}
\end{array},
\end{equation}
which distinguishes the two symmetries by the ancilla action $u_g^a$. For a strong symmetry, the ancilla is $g$-charge neutral, i.e., $u_g^a=\mathbb{1}$. For a weak symmetry, no neutrality condition is imposed, and the ancilla may transform nontrivially ($u_g^a\neq \mathbb{1}$). Clearly, weak symmetry is less restrictive, and consequently, results derived for the weak-symmetry setting also apply, as a special case, to strong symmetries.

Moreover, the symmetry relation in Eq.~\eqref{eq:injectivity_weak} ensures that both the quantized topological responses and the string-order selection rule remain robust for symmetry-protected mixed states~\cite{huang2025prl_mspt}. This follows as in the pure-state case: The symmetry-twisted transfer matrix
\begin{equation} 
T(g_2)\equiv \begin{array}{c} \begin{tikzpicture}[scale=.45, baseline={([yshift=0ex]current bounding box.center)}, thick] \ATensorLPDO{0.4,0}{$A$} \AdagTensorLPDO{0.4,2.8}{$A^*$} \wTensor{-0.2,0.9} {$u_{g_2}$} \draw[red](1,0.7)--(1,2); \end{tikzpicture} \end{array}, \label{eq:transfer_matrix} 
\end{equation}
commutes with the flux operator $V_{g_1}^* \otimes V_{g_1}$.
Therefore, in the large-system limit, the leading eigenvector of $T(g_2)$ which gives the dominant contribution, remains in a one-dimensional representation of $g_1$, and the preceding pure-state derivation carries over directly. 
For a strong $g_2$ symmetry, the push-through identity ensures that $T(g_2)$ is isospectral to $T(\mathbb{1})$, so a change in the topological response requires divergent correlation lengths in both the untwisted and twisted ensembles, similar to the pure state.
Nevertheless, the mixed-state case can differ crucially from the pure-state case. In particular, for weak $g_2$ symmetry, $u_{g_2}$ cannot generally be pushed onto the virtual bond, so $T(g_2)$ need not be isospectral for different $g_2$. Consequently, both the symmetry gap $\Delta(g_2)$ and the symmetry-twisted partition function $Z_{g_2}$ can acquire a nontrivial $g_2$ dependence.

\section{Hidden Long-Range Correlations at Topological Transitions \label{sec:result}}
This $g_2$ dependence enables a phase-transition scenario impossible in pure states: the symmetry-twisted ensemble can become singular while the untwisted one retains a finite correlation length. For pure states, the symmetry gap is independent of $g_2$, so its closing necessarily coincides with a divergent conventional correlation length. In weakly symmetric mixed states, however, the twisted and untwisted gaps can close independently, from which we further identify an isolated closing of the symmetry gap as a distinct mechanism for mixed-state topological transitions: the topological response can change while conventional correlations remain short-ranged, with critical behavior appearing only in the symmetry-twisted ensemble.

To expose the associated singular behavior, we consider expectation values in the symmetry-twisted ensemble [Eq.~\eqref{eq:expectation_twisted_ensemble}, reproduced here for readers' convenience],
\begin{equation}
\langle \dots \rangle_{g_2}
\equiv
\frac{\operatorname{Tr}\left[\rho U_{g_2}(\dots)\right]}
{\operatorname{Tr}\left(\rho U_{g_2}\right)},
\end{equation}
and establish the following results:

\begin{proposition}
Let $p$ be a model parameter, and suppose the topological response changes from
$e^{i\mathcal{Q}_{<}(g_1,g_2)}$ at $p< p_c$ to
$e^{i\mathcal{Q}_>(g_1,g_2)}$
at $p>p_c$, where
$e^{i\mathcal{Q}_{>}(g_1,g_2)}\neq e^{i\mathcal{Q}_{<}(g_1,g_2)}$. Then, at $p =p_c$, 
\begin{enumerate}
    \item The $g_2$-symmetry gap closes;
    \item The $g_2$-twisted ensemble exhibits long-range correlations, i.e., $\lim_{|i-j|\gg 1}
\left|\langle O_iO_j\rangle_{g_2}-\langle O_i\rangle_{g_2}\langle O_j\rangle_{g_2}\right|
\neq 0$. Here, $O_i$ is a suitable local operator at site $i$ that transforms as
$ U_{g_1}O_iU_{g_1}^\dagger=e^{i\Delta\mathcal{Q}(g_1,g_2)}O_i$ with $\Delta\mathcal{Q}(g_1,g_2)\equiv \mathcal{Q}_>(g_1,g_2)-\mathcal{Q}_<(g_1,g_2)$.
\end{enumerate}

\end{proposition}
Statement~1, the closing of the symmetry gap, follows directly from the large-system expression for the mixed-state topological response, which is dominated by the leading eigenvector of the symmetry-twisted transfer matrix $T(g_2)$. Namely, the mixed-state counterpart of Eq.~\eqref{eq:topological_response_pure} at $p\neq p_c$ is
\begin{equation}
\label{eq:topological_response_mixed_simple}
e^{i\mathcal{Q}(g_1,g_2)}
=
\begin{array}{c}
\begin{tikzpicture}[scale=.45, baseline={([yshift=0ex]current bounding box.center)}, thick]




\xTensor{2.5,2.8}{$V_{g_1}^{*}$}
\xTensor{2.5,0}{$V_{g_1}$}

\draw[rounded corners] (1.5,2.8)--(0.95,2.8)--(0.95,0)--(1.5,0);
\filldraw[fill=vector] (0.45,1.0) rectangle (1.45,1.8);
\draw (0.95,1.4) node {\scriptsize $L_{g_2}^{0}$};

\draw[rounded corners] (3.5,2.8)--(4.05,2.8)--(4.05,0)--(3.5,0);
\filldraw[fill=vector] (3.55,1.0) rectangle (4.55,1.8);
\draw (4.05,1.4) node {\scriptsize $R_{g_2}^{0}$};
\end{tikzpicture}.
\end{array}
\end{equation}
Here, $|R_{g_2}^0)$ and $(L_{g_2}^0|$ are the leading right (left) eigenvector of $T(g_2)$, and thus the topological response is given by the eigenvalue of the flux operator associated with the leading eigenstate of $T(g_2)$. Consequently, the topological response can change only when the $g_2$-symmetry gap closes, because such a change requires a switch in the leading eigenstate, implying that the competing leading eigenstates become degenerate at the transition point.

Statement 2 follows from the established symmetry-gap closing (statement 1). For concreteness, we assume that the leading spectral sector of $T(g_2)$
contains a finite number $n_l$ of eigenvalues degenerate in magnitude, i.e.,
\begin{equation}
|\lambda_{g_2}^{(0)}|=|\lambda_{g_2}^{(1)}|=\cdots=|\lambda_{g_2}^{(n_l-1)}|,
\end{equation}
with $n_l\geq 2$, corresponding to a closing of the $g_2$-symmetry gap.
Then, the expectation value of $O_i$, i.e., $\langle O_i \rangle_{g_2}$, vanishes within the leading-eigenvector subspace, i.e.,
\begin{equation}\label{eq:one_point_O}
0=
\begin{array}{c}
\begin{tikzpicture}[
    scale=.45,
    baseline={([yshift=0ex]current bounding box.center)},
    thick
]

\ATensorLPDO{0.4,0}{$A$}
\AdagTensorLPDO{0.4,4.2}{$A^*$}

\wTensor{-0.2,0.9}{$u_{g_2}$}
\wTensor{-0.2,2.3}{$O_i$}

\draw[red] (1,0.7) -- (1,3.5);

\draw[rounded corners]
    (-1.2,4.2) -- (-1.75,4.2)
    -- (-1.75,0) -- (-1.2,0);

\filldraw[fill=vector]
    (-2.25,1.6) rectangle (-1.25,2.6);

\node at (-1.75,2.1)
    {\scriptsize $L_{g_2}^{a}$};

\draw[rounded corners]
    (2.0,4.2) -- (2.55,4.2)
    -- (2.55,0) -- (2.0,0);

\filldraw[fill=vector]
    (2.05,1.6) rectangle (3.05,2.6);

\node at (2.55,2.1)
    {\scriptsize $R_{g_2}^{a}$};

\end{tikzpicture}
\end{array}, a=0, 1, \dots n_l-1.
\end{equation}
This is because $O_i$ is charged under $U_{g_1}$: Given that the right leading eigenvector $|R_{g_2}^a)$ carries flux charge $\mathcal{Q}_a(g_1, g_2)$, 
\begin{equation}
\begin{array}{c}
\begin{tikzpicture}[
    scale=.45,
    baseline={([yshift=0ex]current bounding box.center)},
    thick
]

\draw[rounded corners]
    (2.0,4.2) -- (2.55,4.2)
    -- (2.55,0) -- (2.0,0);

\filldraw[fill=vector]
    (2.05,1.6) rectangle (3.05,2.6);

\node at (2.55,2.1)
    {\scriptsize $R_{g_2}^{a}$};

\xTensor{1, 4.2}{$V_{g_1}^*$}
\xTensor{1, 0}{$V_{g_1}$}

\end{tikzpicture}
\end{array}=e^{i\mathcal{Q}_a(g_1, g_2)}\begin{array}{c}
\begin{tikzpicture}[
    scale=.45,
    baseline={([yshift=0ex]current bounding box.center)},
    thick
]

\draw[rounded corners]
    (2.0,4.2) -- (2.55,4.2)
    -- (2.55,0) -- (2.0,0);

\filldraw[fill=vector]
    (2.05,1.6) rectangle (3.05,2.6);

\node at (2.55,2.1)
    {\scriptsize $R_{g_2}^{a}$};


\end{tikzpicture}
\end{array},
\end{equation}
the corresponding left eigenvector in the presence of $O_i$ carries a charge different by $\Delta\mathcal{Q}(g_1, g_2)$, i.e., 
\begin{equation}
\begin{array}{c}
\begin{tikzpicture}[
    scale=.45,
    baseline={([yshift=0ex]current bounding box.center)},
    thick
]

\ATensorLPDO{0.4,0}{$A$}
\AdagTensorLPDO{0.4,4.2}{$A^*$}

\wTensor{-0.2,0.9}{$u_{g_2}$}
\wTensor{-0.2,2.3}{$O_i$}

\draw[red] (1,0.7) -- (1,3.5);

\draw[rounded corners]
    (-1.2,4.2) -- (-1.75,4.2)
    -- (-1.75,0) -- (-1.2,0);

\filldraw[fill=vector]
    (-2.25,1.6) rectangle (-1.25,2.6);

\node at (-1.75,2.1)
    {\scriptsize $L_{g_2}^{a}$};

\xTensor{2.55, 4.2}{$V_{g_1}^*$}
\xTensor{2.55, 0}{$V_{g_1}$}

\end{tikzpicture}
\end{array}=e^{i\mathcal{Q}^\prime_a(g_1, g_2)}\begin{array}{c}
\begin{tikzpicture}[
    scale=.45,
    baseline={([yshift=0ex]current bounding box.center)},
    thick
]

\ATensorLPDO{0.4,0}{$A$}
\AdagTensorLPDO{0.4,4.2}{$A^*$}

\wTensor{-0.2,0.9}{$u_{g_2}$}
\wTensor{-0.2,2.3}{$O_i$}

\draw[red] (1,0.7) -- (1,3.5);

\draw[rounded corners]
    (-1.2,4.2) -- (-1.75,4.2)
    -- (-1.75,0) -- (-1.2,0);

\filldraw[fill=vector]
    (-2.25,1.6) rectangle (-1.25,2.6);

\node at (-1.75,2.1)
    {\scriptsize $L_{g_2}^{a}$};


\end{tikzpicture}
\end{array},
\label{eq:Oi-leading-contraction}
\end{equation}
with $\mathcal{Q}^\prime_a(g_1, g_2)=[\mathcal{Q}_a(g_1, g_2)+\Delta\mathcal{Q}(g_1, g_2)]$. Thus, the overlap in Eq.~\eqref{eq:one_point_O} vanishes whenever $\Delta\mathcal{Q}(g_1,g_2)\neq 0$.
Moreover, these distinct charges impose selection rules between different leading eigenvectors of $T(g_2)$, thereby giving rise to long-range correlations. Specifically, retaining only contributions from the leading-eigenvector sector, the two-point correlation function for a diagonalizable $T(g_2)$ reduces to
\begin{eqnarray}
&&\langle O_i O_j\rangle_{g_2}\nonumber\\
&=&\sum_{a_{1,2}=0}^{n_l-1}\mathcal{N}_{a_1 a_2}^{-1}
\begin{array}{c}
\begin{tikzpicture}[
    scale=.45,
    baseline={([yshift=0ex]current bounding box.center)},
    thick
]

\ATensorLPDO{0.4,0}{$A$}
\AdagTensorLPDO{0.4,4.2}{$A^*$}

\wTensor{-0.2,0.9}{$u_{g_2}$}
\wTensor{-0.2,2.3}{$O_i$}

\draw[red] (1,0.7) -- (1,3.5);

\draw[rounded corners]
    (-1.2,4.2) -- (-1.75,4.2)
    -- (-1.75,0) -- (-1.2,0);

\filldraw[fill=vector]
    (-2.25,1.6) rectangle (-1.25,2.6);

\node at (-1.75,2.1)
    {\scriptsize $L_{g_2}^{a_1}$};

\draw[rounded corners]
    (2.0,4.2) -- (2.55,4.2)
    -- (2.55,0) -- (2.0,0);

\filldraw[fill=vector]
    (2.05,1.6) rectangle (3.05,2.6);

\node at (2.55,2.1)
    {\scriptsize $R_{g_2}^{a_2}$};

\end{tikzpicture}
\end{array}\times 
\begin{array}{c}
\begin{tikzpicture}[
    scale=.45,
    baseline={([yshift=0ex]current bounding box.center)},
    thick
]

\ATensorLPDO{0.4,0}{$A$}
\AdagTensorLPDO{0.4,4.2}{$A^*$}

\wTensor{-0.2,0.9}{$u_{g_2}$}
\wTensor{-0.2,2.3}{$O_j$}

\draw[red] (1,0.7) -- (1,3.5);

\draw[rounded corners]
    (-1.2,4.2) -- (-1.75,4.2)
    -- (-1.75,0) -- (-1.2,0);

\filldraw[fill=vector]
    (-2.25,1.6) rectangle (-1.25,2.6);

\node at (-1.75,2.1)
    {\scriptsize $L_{g_2}^{a_2}$};

\draw[rounded corners]
    (2.0,4.2) -- (2.55,4.2)
    -- (2.55,0) -- (2.0,0);

\filldraw[fill=vector]
    (2.05,1.6) rectangle (3.05,2.6);

\node at (2.55,2.1)
    {\scriptsize $R_{g_2}^{a_1}$};

\end{tikzpicture}
\end{array},\nonumber\\
\label{eq:Oi-leading-contraction}
\end{eqnarray}
where $\mathcal{N}_{a_1a_2}$ is a normalization factor, i.e., 
\begin{equation}
\mathcal{N}_{a_1 a_2}=\frac{(\lambda_{g_2}^{(a_2)})^{|i-j|-1}(\lambda_{g_2}^{(a_1)})^{N-(|i-j|+1)}}{\sum_{a=0}^{n_l-1}(\lambda_{g_2}^{(a)})^N}\sim \mathcal{O}(1),~\label{eq:N_statement2}
\end{equation}
where we have assumed the absence of fine-tuned cancellations among the leading-eigenvalue contributions, such that the normalization factor remains $\mathcal{O}(1)$.
The change in the topological response  $\mathcal{Q}(g_1,g_2)=
\begin{cases}
\mathcal{Q}_<(g_1,g_2), & p<p_c\\
\mathcal{Q}_>(g_1,g_2), & p>p_c
\end{cases}$, necessitates the existence of a pair of leading eigenvectors $|R_{g_2}^{a})$ and $( L_{g_2}^{a'}|$ whose $(V_{g_1}^{*}\otimes V_{g_1})$ eigenvalues differ by $\Delta\mathcal{Q}$. Thus, the corresponding matrix element is of order one in the absence of fine tuning, i.e.,
\begin{equation}
\mathcal{Q}_{a_2}=\mathcal{Q}_{a_1}+\Delta\mathcal{Q}:\ 
\begin{array}{c}
\begin{tikzpicture}[
    scale=.45,
    baseline={([yshift=0ex]current bounding box.center)},
    thick
]

\ATensorLPDO{0.4,0}{$A$}
\AdagTensorLPDO{0.4,4.2}{$A^*$}

\wTensor{-0.2,0.9}{$u_{g_2}$}
\wTensor{-0.2,2.3}{$O_i$}

\draw[red] (1,0.7) -- (1,3.5);

\draw[rounded corners]
    (-1.2,4.2) -- (-1.75,4.2)
    -- (-1.75,0) -- (-1.2,0);

\filldraw[fill=vector]
    (-2.25,1.6) rectangle (-1.25,2.6);

\node at (-1.75,2.1)
    {\scriptsize $L_{g_2}^{a_1}$};

\draw[rounded corners]
    (2.0,4.2) -- (2.55,4.2)
    -- (2.55,0) -- (2.0,0);

\filldraw[fill=vector]
    (2.05,1.6) rectangle (3.05,2.6);

\node at (2.55,2.1)
    {\scriptsize $R_{g_2}^{a_2}$};

\end{tikzpicture}
\end{array}\sim \mathcal{O}(1),
\end{equation}
whereas all other matrix elements vanish.
Taken together, for a
diagonalizable $T(g_2)$, these results show that $\langle O_i O_j\rangle_{g_2}$ remains finite even in the limit $|i-j|\gg 1$.
For a non-diagonalizable leading block, the nilpotent part of the
Jordan decomposition can produce additional polynomial prefactors in the
numerator of Eq.~\eqref{eq:Oi-leading-contraction}, yielding additional
nondecaying contributions without altering the conclusion of long-range correlations.
\\

Below, we also argue that:
\begin{proposition}
In the large-system limit, the symmetry-twisted partition function develops a singularity at the transition point, manifested as
\begin{equation}
p=p_c,\qquad \frac{1}{N}\partial_p\ln |\operatorname{Tr}(\rho U_{g_2})|\ \mathrm{singular}.
\end{equation}
\end{proposition}

This follows because, in the large-system limit and away from the transition point $p=p_c$, $|\operatorname{Tr}(\rho U_{g_2})|=|\lambda_{g_2}^{(0)}|^N\left[1+\mathcal{O}\left(|\lambda_{g_2}^{(1)}/\lambda_{g_2}^{(0)}|^N\right)\right]$. Thus, as $p$ is tuned through the transition, the dominant eigenvalue switches, producing a cusp in $\ln|\operatorname{Tr}(\rho U_{g_2})|$. 

The mechanism underlying these singularities is closely related to the
pure-state MPS quantum phase transitions~\cite{wolf2006prl}.
There, a crossing of dominant transfer-matrix eigenvalues gives rise to a divergent correlation length and nonanalytic observables, while the correlations become nondecaying rather than exhibiting the algebraic decay characteristic of conventional continuous criticality. The same spectral mechanism operates here in the
symmetry-twisted transfer matrix. Crucially, however, for weakly symmetric
mixed states the singularity occurs in the twisted sector,
while the untwisted sector remains gapped and conventional correlations remain short-ranged.
Finally, related singular signatures have also been observed in reduced
density matrices of SPT states~\cite{huang2025prl,huang2025prl_mspt,xu2025arxiv}
and in finite-temperature spin-$1/2$ chains~\cite{zang2024prl,zhou2024nsr}.
In those settings, the singular behavior originates from a mixed-state
anomaly~\cite{huang2022prb,kawabata2024prl,zang2024prl,xu2025arxiv,
lessa2025prx,zhou2024nsr,huang2026arxiv}, such as an SPT boundary anomaly
or a Lieb-Schultz-Mattis anomaly. By contrast, here the singularity is
generated by a decoherence-driven topological transition.

\section{Example 1: Transition from a cluster state to a product state \label{sec:ex1_cluster}}
To illustrate the physics implied by these results, we turn to an exactly solvable model: an even-site cluster state decohering into the trivial product state $|+\rangle^{\otimes N}$, where $|+\rangle$ is the $+1$ eigenstate of the Pauli $X$ operator. We also investigate its detection with a quantum microscope~\cite{gross2017science} by Monte Carlo sampling site-resolved measurement snapshots.

\paragraph{Model.--} The cluster state is an SPT state of spin-$\frac{1}{2}$ degrees of freedom, protected by a $\mathbb{Z}_2\times\mathbb{Z}_2$ symmetry generated by
\begin{equation}
S_{\mathrm e}\equiv \prod_{i\in 2\mathbb{Z}} X_i,
\qquad
S_{\mathrm o}\equiv \prod_{i\in 2\mathbb{Z}+1} X_i,
\end{equation}
which act on the even and odd sublattices, respectively. In the MPS representation, the $A$-tensors are
\begin{equation}
A^{++}=\frac{1}{2}\left(\begin{array}{cc}
1 & 0\\
0 & 1
\end{array}\right),\ A^{+-}=\frac{1}{2}\left(\begin{array}{cc}
1 & 0\\
0 & -1
\end{array}\right),
\end{equation}
and
\begin{equation}
\ A^{-+}=\frac{1}{2}\left(\begin{array}{cc}
0 & 1\\
1 & 0
\end{array}\right),\ \text{and}\ A^{--}=\frac{1}{2}\left(\begin{array}{cc}
0 & -1\\
1 & 0
\end{array}\right),
\end{equation}
where the first (second) superscript labels the odd (even) site, and the two signs label the $X$-eigenstates of the corresponding physical sites. 
Correspondingly, the local actions of $S_{\mathrm{e, o}}$ [Eq.~\eqref{eq:injectivity_pure}] are
 \begin{equation}
\begin{tikzpicture}[scale=.4, baseline={([yshift=-1.3ex]current bounding box.center)}, thick]
    \ATensorMPS{0,0}{$A$}
    \wTensor{0,0.9}{$X\mathbb{1}$}  
  \end{tikzpicture}=
\begin{tikzpicture}[scale=.4, baseline={([yshift=-1.3ex]current bounding box.center)}, thick]
    \ATensorMPS{0,0}{$A$}
     \xTensor{1.6, 0}{$Z$}
     \xTensor{-1.6, 0}{$Z$}
  \end{tikzpicture},
\end{equation}
and 
\begin{equation}
\begin{tikzpicture}[scale=.4, baseline={([yshift=-1.3ex]current bounding box.center)}, thick]
    \ATensorMPS{0,0}{$A$}
    \wTensor{0,0.9}{$\mathbb{1}X$}  
  \end{tikzpicture}=
\begin{tikzpicture}[scale=.4, baseline={([yshift=-1.3ex]current bounding box.center)}, thick]
    \ATensorMPS{0,0}{$A$}
     \xTensor{1.6, 0}{$X$}
     \xTensor{-1.6, 0}{$X$}
  \end{tikzpicture},
\end{equation}
where $Z$ is the Pauli-$Z$ matrix, $\mathbb{1}X\equiv \mathbb{1}\otimes X$, and similarly for the other operator products. 
The nontrivial SPT character of this state, as discussed above, is encoded in its quantized responses to symmetry-flux insertion, 
\begin{equation}
e^{i\mathcal{Q}(S_{\mathrm{e}},S_{\mathrm{o}})}= e^{i\mathcal{Q}(S_{\mathrm{o}},S_{\mathrm{e}})}=-1.
\end{equation}

We then drive a topological transition to the trivial product state by applying a local decoherence channel,

\begin{equation}
\rho(p)=\mathcal{E}_1\circ \mathcal{E}_2\dots \circ\mathcal{E}_N(
|\psi_{\mathrm{cluster}}\rangle
\langle\psi_{\mathrm{cluster}}|),
\end{equation}
where $\mathcal{E}_i(\cdot)\equiv\sum_\alpha K_{\alpha, i}(\cdot)K_{\alpha, i}^\dagger$, with local weakly symmetric Kraus operators at site $i$ given by,
\begin{equation} 
\begin{cases} 
K_{0,\ i}=P_{i}^{+}+\sqrt{1-p}P_{i}^{-}\\ K_{1,\ i}=\sqrt{p}Z_{i}P_{i}^{-} 
\end{cases}, \mathrm{with}\ p\in [0, 1]. \end{equation}
Here, $P_i^{\pm}$ project onto the $\pm1$ eigenstates of $X_i$. Physically, the local operator $K_{0,i}$ assigns different weights to the states $|\pm\rangle$, interpolating from equal weighting at $p=0$ to a bias toward $|+\rangle$ at $p=1$. Meanwhile, $K_{1,i}$ has no effect at $p=0$ but flips the $|-\rangle$ state at $p=1$. The combined action of the channel therefore leaves the initial state unchanged at $p=0$ and maps it to the trivial product state $|+\rangle^{\otimes N}$ at $p=1$, i.e.,  (1) $p=0$: $K_{0,i}=\mathbb{1},\ K_{1,i}=0$, and thus $e^{i\mathcal{Q}(S_{\mathrm{e}},S_{\mathrm{o}})}= e^{i\mathcal{Q}(S_{\mathrm{o}},S_{\mathrm{e}})}=-1$; (2) $p=1$: $K_{0,i}=P^{+}, K_{1,i}=Z_{i}P_{i}^{-}$, and thus $e^{i\mathcal{Q}(S_{\mathrm{e}},S_{\mathrm{o}})}= e^{i\mathcal{Q}(S_{\mathrm{o}},S_{\mathrm{e}})}=1$. Moreover, the channel exhibits an exact $p\leftrightarrow 1-p$ duality, which pins a unique symmetry-gap closing to $p_c=1/2$, as confirmed below. To see this, its adjoint action reads
\begin{equation}
\mathcal{E}_i^\dagger(X_i)|_p
=\sum_\alpha K_{\alpha,i}^\dagger X_i K_{\alpha,i}
=P_i^+ +(2p-1)P_i^-,
\end{equation}
which by using $X_i=P_i^+-P_i^-$, leads to
\begin{equation}
\mathcal{E}_i^\dagger(X_i)|_p
=X_i\mathcal{E}_i^\dagger(X_i)|_{1-p}.\label{eq:dual_channel}
\end{equation}
The twisted transfer matrices $T(S_{\mathrm{o}})|_p$ and $T(S_{\mathrm{o}})|_{1-p}$ are thus related by a unitary similarity transformation in the virtual space, up to an overall phase, by using the push-through identity. Consequently, their spectra have identical eigenvalue magnitudes, and the symmetry gap is symmetric about $p=1/2$.

\paragraph{Symmetry gap and divergent correlation length.--} We now test the results of Sec.~\ref{sec:result} by computing the spectrum of the symmetry-twisted transfer matrix and the correlation length in the corresponding twisted ensemble. Taking $g_2=S_{\mathrm{o}}$, the transfer matrix $T(g_2)$ can be diagonalized as,
\begin{eqnarray}
T\left(S_{\mathrm{o}}\right)&=&\left(1-p\right)|\psi_{1}\rangle\langle\psi_{1}|+p|\psi_{2}\rangle\langle\psi_{2}|\nonumber\\
&&+0\times\left(|\psi_{3}\rangle\langle\psi_{3}|+|\psi_{4}\rangle\langle\psi_{4}|\right),
\end{eqnarray}
with eigenvectors,
\begin{equation}
|\psi_{1}\rangle=\frac{1}{\sqrt{2}}\left(\begin{array}{c}
-1\\
0\\
0\\
1
\end{array}\right), \ |\psi_{2}\rangle=\frac{1}{\sqrt{2}}\left(\begin{array}{c}
1\\
0\\
0\\
1
\end{array}\right),\
\end{equation}
and
\begin{equation}
 |\psi_{3}\rangle=\frac{1}{\sqrt{2}}\left(\begin{array}{c}
0\\
-1\\
1\\
0
\end{array}\right),\ |\psi_{4}\rangle=\frac{1}{\sqrt{2}}\left(\begin{array}{c}
0\\
1\\
1\\
0
\end{array}\right).
\end{equation}
The symmetry gap is thus given as,
\begin{equation}
\Delta(S_{\mathrm{o}})= |1-2p|,  
\end{equation}
which closes at $p_c=\frac{1}{2}$, consistent with the duality relation Eq.~\eqref{eq:dual_channel}. Also, the correlation length of the twisted ensemble diverges at the same transition point. To demonstrate this, we evaluate the following correlation function between even sites,
\begin{equation}
\langle Z_{2i}Z_{2j}\rangle_{S_\mathrm{o}}-\langle Z_{2i}\rangle_{S_\mathrm{o}}\langle Z_{2j}\rangle_{S_\mathrm{o}},
\end{equation}
where $Z_{2i}$ is odd under $S_{\mathrm{e}}$, i.e.,
$S_{\mathrm{e}}Z_{2i}S_{\mathrm{e}}^\dagger=-Z_{2i}$.
The resulting correlation function, in the limit $|i-j|\gg 1$, is
\begin{eqnarray}
 &  & \langle Z_{2i}Z_{2j}\rangle_{S_{\mathrm{o}}}-\langle Z_{2i}\rangle_{S_{\mathrm{o}}}\langle Z_{2j}\rangle_{S_{\mathrm{o}}}\nonumber \\
 & \overset{\left|i-j\right|\gg1}{=} & \begin{cases}
\left(1-p\right)\left(\frac{p}{1-p}\right)^{\left|i-j\right|} & p<p_{c}\\
\frac{1}{2} & p=p_{c}\\
(1-p)\left(\frac{1-p}{p}\right)^{\left|i-j\right|} & p>p_{c}
\end{cases},
\end{eqnarray}
from which we can infer a divergent correlation length at the transition point, 
\begin{equation}
\xi(g)=\begin{cases}
-\frac{2}{\ln\left|\frac{p}{1-p}\right|} & p<p_{c}\\
\infty & p=p_{c}\\
-\frac{2}{\ln\left|\frac{1-p}{p}\right|} & p>p_{c}
\end{cases}, 
\end{equation}
with $g=S_o$. This can be related to the symmetry gap by expanding in the distance from criticality, $|p-p_c|$, 
\begin{equation}
|p-p_c|\ll 1:\ \xi(g)\sim \frac{1}{2|p-p_c|}= \frac{1}{\Delta(g)}.
\end{equation}
Finally, a cusp in the symmetry-twisted partition function is also observed at the transition point in the thermodynamic limit, i.e.,
\begin{equation}
-\ln \left|\operatorname{Tr}(\rho S_o)\right|/N=
\begin{cases}
-\frac{1}{2}\ln(1-p), & p<p_c\\
-\frac{1}{2}\ln p, & p>p_c
\end{cases},
\end{equation}
in agreement with the results of Sec.~\ref{sec:result}.
\begin{figure}
\includegraphics[width=1\linewidth,keepaspectratio]{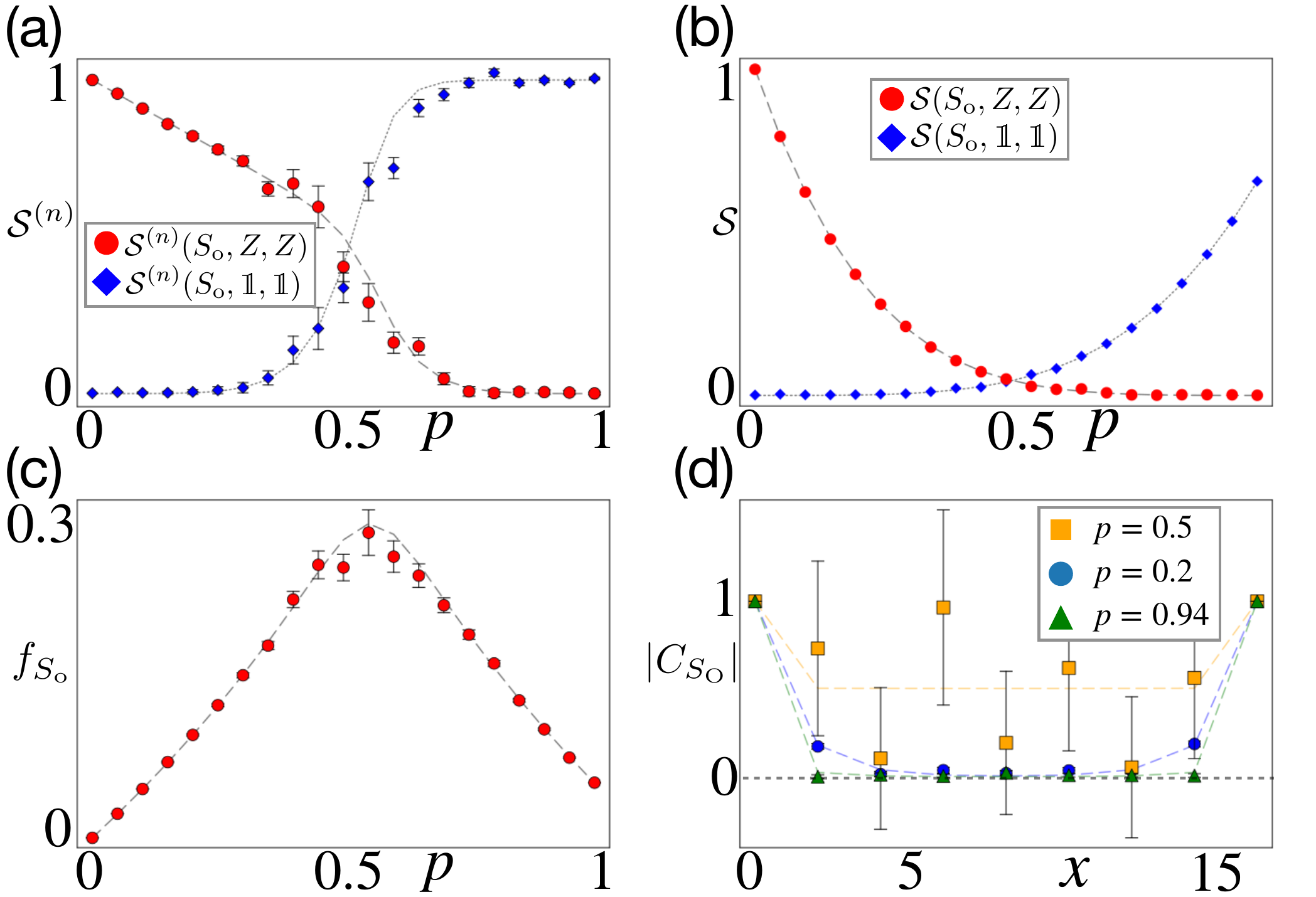}\caption{Numerically simulated experimental data for a periodic 16-site chain, with error bars estimated from measurement snapshots: (a) the normalized string order parameters $\mathcal{S}^{(n)}(S_{\mathrm{o}},Z,Z)$ and $\mathcal{S}^{(n)}(S_{\mathrm{o}},\mathbb{1},\mathbb{1})$, with the end points at $i=4, 12$; (b) the corresponding bare string order parameter $\mathcal{S}$; (c) the symmetry-twisted free-energy density, $f_{S_{\mathrm{O}}}\equiv-\frac{1}{N}\ln\left|\operatorname{Tr}(\rho S_{\mathrm{o}})\right|$; and (d) the twisted correlation function, $|C_{S_{\mathrm{O}}}(x)|\equiv|\langle Z_2 Z_{2+x}\rangle_{S_{\mathrm{o}}}| $. Dashed lines denote direct tensor-network calculations, while markers indicate Monte Carlo estimates obtained from $10^5$ simulated quantum-microscope snapshots sampled from the MPO density matrix.
\label{fig:cluster_snapshot}}
\end{figure}

We also note that, for the onsite decoherence channels considered here, the untwisted transfer matrix remains identical to that of the initial pure state due to trace preservation. By contrast, the symmetry-twisted transfer matrix acquires a nontrivial dependence on the channel through the dressed symmetry operator $\mathcal{E}_p^\dagger(u_g)$.

\paragraph{Measurement via quantum microscope.--} In practice, the string-order parameter, symmetry-twisted partition function, and twisted correlation function can be extracted directly from site- or qubit-resolved projective-measurement snapshots. Such measurements are available on several current quantum-simulation platforms, including quantum gas microscopes~\cite{gross2017science}, programmable Rydberg-atom arrays~\cite{semeghini2021science}, trapped-ion simulators~\cite{zhang2017nature}, and superconducting-qubit processors~\cite{satzinger2021science}. 
In particular, resolving these nonlocal observables at fixed relative
accuracy may require exponentially many measurement shots, particularly near
the transition. We therefore focus on mesoscopic system sizes, consistent with
existing experiments and proposals
~\cite{bardyn2018prx,sompet2022nature,semeghini2021science,
zhang2017nature,satzinger2021science}, and simulate this regime using
ITensor~\cite{itensor}.
To this end, we first prepare the pure cluster state as an MPS using the density-matrix renormalization group (DMRG)~\cite{white1992prl,white1993prb} and construct the decohered density matrix as an MPO by applying the local decoherence channel. To measure $Z$ on the even sites and $X$ on the odd sites, we apply a Hadamard rotation to each odd site and draw projective-measurement outcomes from the resulting MPO using the ITensor sampling routine. Each sample yields a site-resolved bit string, which we map to $\pm1$ measurement outcomes. The observables are thus evaluated directly from each snapshot and averaged over the sampled ensemble.

The measured normalized string-order parameter obeys the expected selection rules, as shown in Fig.~\ref{fig:cluster_snapshot}(a), while the bare values are shown for comparison in Fig.~\ref{fig:cluster_snapshot}(b). The measured symmetry-twisted partition function is shown in Fig.~\ref{fig:cluster_snapshot}(c); it becomes relatively noisy near the transition point. Correspondingly, the twisted correlation function in Fig.~\ref{fig:cluster_snapshot}(d) exhibits strong fluctuations in this regime but decays rapidly away from the transition. 

\begin{figure}
\includegraphics[width=1\linewidth,keepaspectratio]{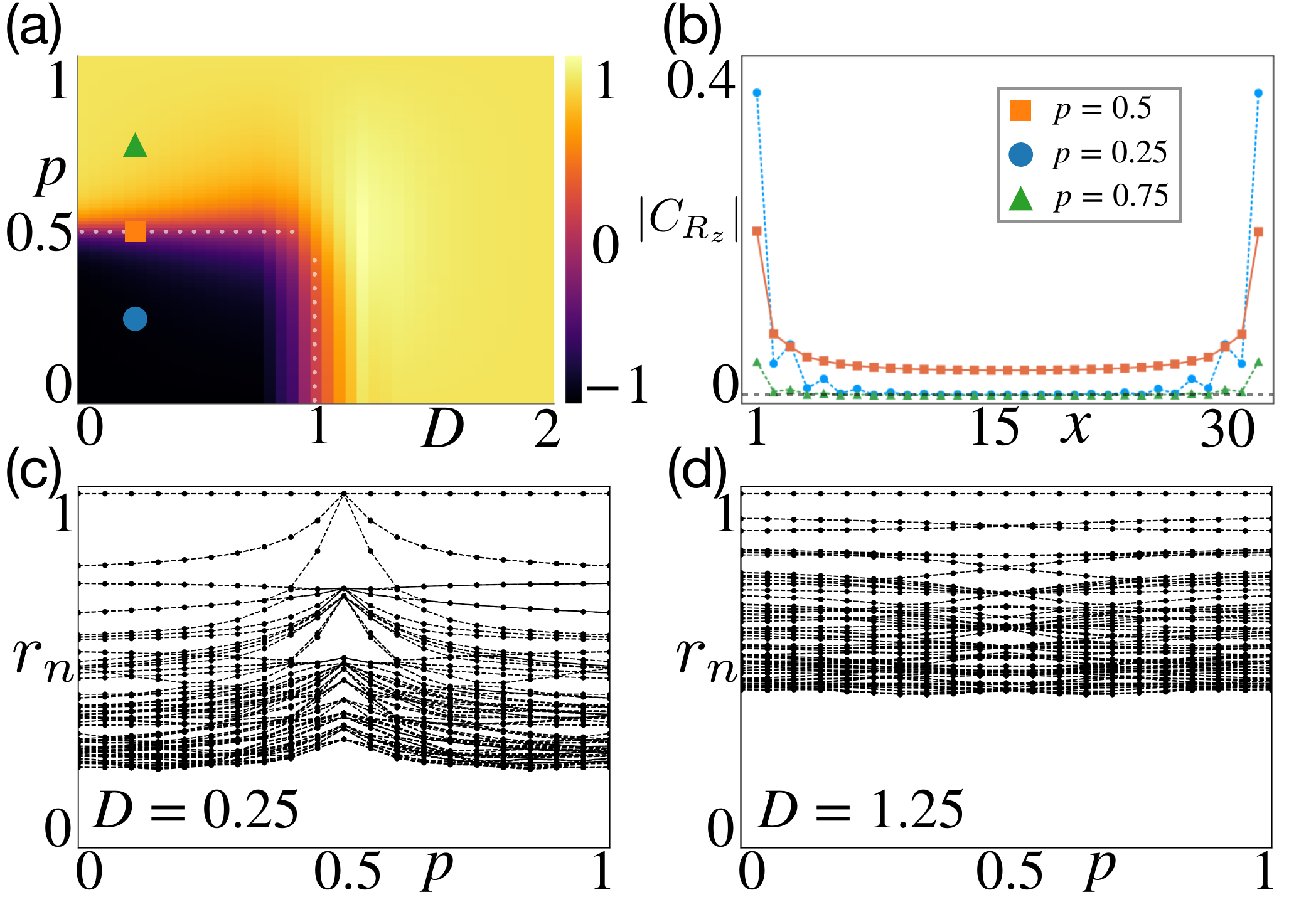}\caption{Numerical results for the 32-site deformed Haldane chain under decoherence. (a) Phase diagram in the $D$-$p$ plane, obtained from the quantized responses as order parameters; the dotted line marks the phase boundary. (b) Symmetry-twisted correlation function ($|C_{R_z}(x)|\equiv |\left\langle S_{z,1}S_{z,1+x}\right\rangle_{R_z}|$) at $D=0.25$ for $p=0.25$, $0.5$, and $0.75$. Nondecaying correlations persist at $p=0.5$, consistent with the phase diagram in (a). (c, d) $r_n\equiv |\lambda_{R_z}^{(n)}/\lambda_{R_z}^{(0)}|$ at $D=0.25, 1.25$, computed from an initial MPS with maximum bond dimension $100$. The symmetry gap closes at $p=0.5$ for the former, but remains open for the latter.
\label{fig:dHaldane}}
\end{figure}
\section{Example 2: From a spin-1 topological ground state to a trivial product state\label{sec:ex2_dHaldane}}
Moving beyond the cluster state, which admits an exact MPS representation, we test our results on the ground state of a generic spin-1 chain.

Consider the spin-1 Heisenberg chain with an on-site anisotropy,
\begin{equation}
H=\sum_i \boldsymbol{S}_{i+1}\cdot\boldsymbol{S}_{i}+D\sum_i S_{z,i}^2, 
\end{equation}
where $D>0$ favors the $S_z=0$ state.
Meanwhile,
$\boldsymbol{S}_{i}\equiv (S_{x, i}, S_{y, i}, S_{z,i})$ denotes spin-$1$ operators at site $i$, i.e., 
\begin{equation}
    S_{x,i}=\frac{1}{\sqrt{2}}\left(\begin{array}{ccc}
0 & 1 & 0\\
1 & 0 & 1\\
0 & 1 & 0
\end{array}\right),\ S_{y, i}=\frac{1}{\sqrt{2}}\left(\begin{array}{ccc}
0 & -i & 0\\
i & 0 & -i\\
0 & i & 0
\end{array}\right),
\end{equation}
and 
\begin{equation}
S_{z, i}=\left(\begin{array}{ccc}
1 & 0 & 0\\
0 & 0 & 0\\
0 & 0 & -1
\end{array}\right),
\end{equation}
where the site index is omitted when no confusion can arise.
The model is protected by the $\mathbb{Z}_2\times \mathbb{Z}_2$ symmetry, generated by $R_x\equiv e^{i\pi S_x}$ and $R_z\equiv e^{i\pi S_z}$. Its phase diagram, established in Refs.~\cite{gu2009prb,pollmann2012prb,hu2011prb}, consists of an SPT phase for $D<D_c$ and a topologically trivial phase for $D>D_c$, separated by a quantum critical point at $D_c\simeq 1$~\cite{hu2011prb}.

As with the cluster state, we then apply a decoherence channel that drives the state toward the trivial ($S_z=0$) product state $|0\rangle^{\otimes N}$, i.e.,
\begin{equation}
\rho(p)=\mathcal{E}_1\circ \mathcal{E}_2\dots\circ \mathcal{E}_N(
|\psi_{\mathrm{DH}}\rangle
\langle\psi_{\mathrm{DH}}|),
\end{equation}
where $|\psi_{\mathrm{DH}}\rangle$ denotes the ground state of the deformed Haldane chain. $\mathcal{E}_i(\cdot)=\sum_\alpha K_{\alpha, i}(\cdot)K_{\alpha, i}^\dagger$, with the weakly symmetric local Kraus operators at site $i$ given by,

\begin{equation}
\begin{cases}
K_{0,\ i}=P_{i}^{0}+\sqrt{1-p} P_{i}^{1}+\sqrt{1-p}P_{i}^{-1}\\
K_{1,\ i}=\sqrt{\frac{p}{2}}M^{0, 1}_{i}P_{i}^{1}+\sqrt{\frac{p}{2}} M^{0, -1}_{i}P_{i}^{-1}\\
K_{2,\ i}=\sqrt{\frac{p}{2}}M^{0, 1}_{i}P_{i}^{1}-\sqrt{\frac{p}{2}} M^{0, -1}_{i}P_{i}^{-1}
\end{cases},
\end{equation}
Here, $P_i^{-1}$, $P_i^{0}$, and $P_i^{1}$ project onto the $S^z=-1$, $0$, and $1$ states, respectively. Meanwhile, $M^{0, 1}$ and $M^{0, -1}$ map the $|+1\rangle$ and the $|-1\rangle$ states to the $|0\rangle$ state, respectively, i.e., 
\begin{equation}
M^{0, 1}=\left(\begin{array}{ccc}
0 & 1 & 0\\
1 & 0 & 0\\
0 & 0 & 0
\end{array}\right),\ \text{and}\ 
M^{0, -1}=\left(\begin{array}{ccc}
0 & 0 & 0\\
0 & 0 & 1\\
0 & 1 & 0
\end{array}\right).
\end{equation}
Consequently, the quantum channel acts as an identity in the limit $p=0$, i.e., 
\begin{equation}
K_{0, i}=\mathbb{1}, \mathrm{and}\ K_{1, i}=K_{2, i}=0,
\end{equation}
whereas at $p=1$, it drives the system toward the product state $|0\rangle^{\otimes N}$, as required.

\begin{figure}[t]
\includegraphics[width=0.75\linewidth,
keepaspectratio]{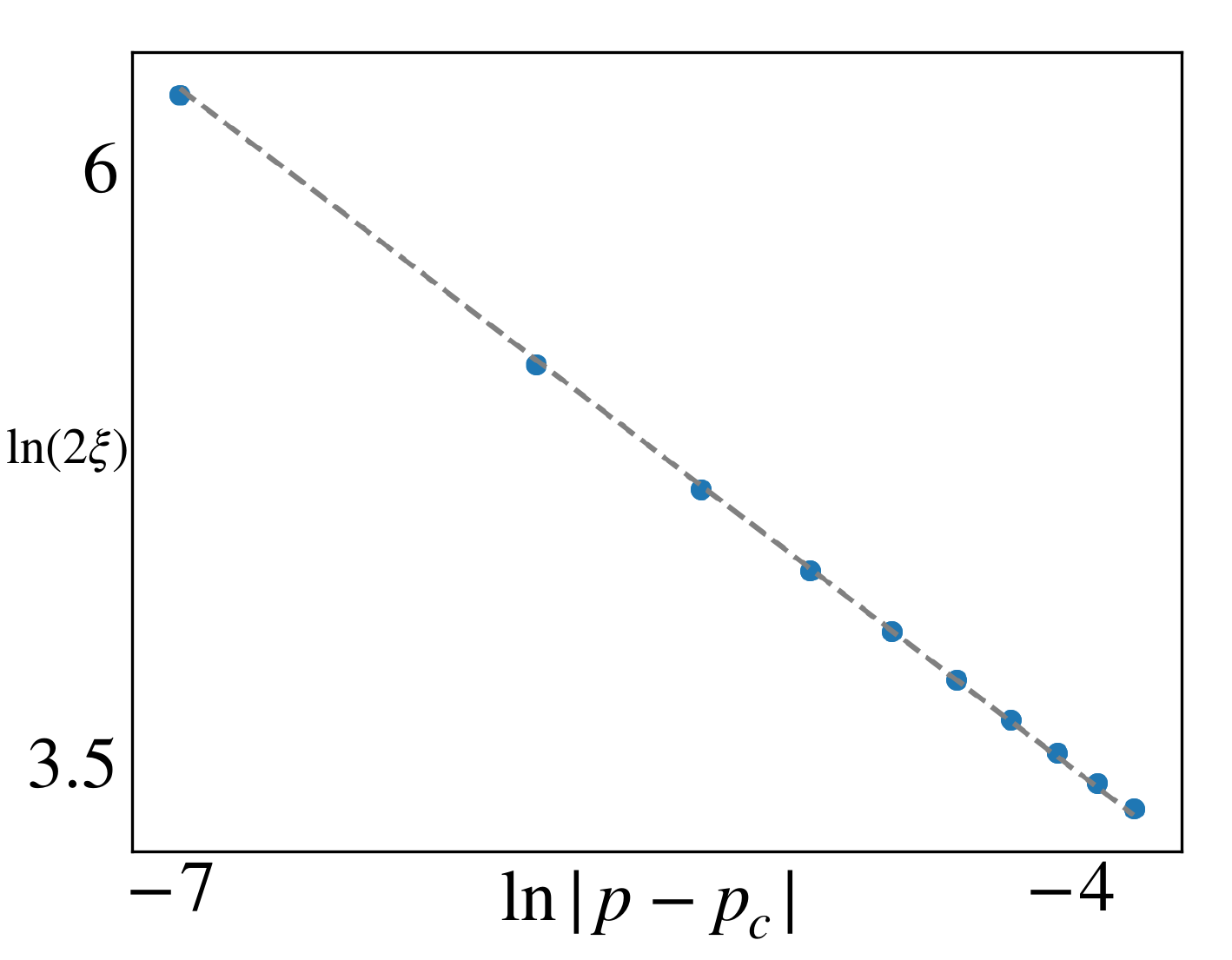}\caption{Numerical results for $\xi(R_z)$ versus $p$, obtained from the transfer-matrix spectrum of an infinite MPS computed using iDMRG with a two-site unit cell and maximum bond dimension $100$ [see Eq.~\eqref{eq:correlation_length}]. The dashed line shows the expected linear dependence of $\ln[2\xi(R_z)]$ on $\ln|p-p_c|$.
\label{fig:xi_fit}}
\end{figure}

We then numerically calculate the phase diagram in the $D$-$p$ plane [Fig.~\ref{fig:dHaldane}(a)], using the quantized response
\begin{equation}
\frac{\operatorname{Tr}\left[\rho(R_x)R_z^{\otimes N}\right]}
{\operatorname{Tr}\left[\rho(\mathbb{1})R_z^{\otimes N}\right]}
\end{equation}
as the order parameter.
To further characterize the phase structure, we examine the symmetry-twisted two-point correlation function,
\begin{equation}
|\left\langle S_{z,1}S_{z,1+x}\right\rangle_{R_z}|,
\end{equation}
where $R_x S_z R_x=-S_z$. As shown in Fig.~\ref{fig:dHaldane}(b), long-range correlations occur at $p=0.5$, consistent with the results of Sec.~\ref{sec:result}. In particular, the twisted correlation length $\xi(R_z)$ near the transition, shown numerically in Fig.~\ref{fig:xi_fit}, scales as
\begin{equation}
\xi(R_z)\sim |p-p_c|^{-1},
\end{equation}
consistent with a linear crossing of the dominant symmetry-twisted
transfer-matrix eigenvalues near the transition.
For comparison, the corresponding Hamiltonian-driven pure-state transition
exhibits
\begin{equation}
\xi(R_z)\sim |D-D_c|^{-1.472}
\end{equation}
as found numerically in Ref.~\cite{hu2011prb}. Finally, Figs.~\ref{fig:dHaldane}(c,d) confirm that the symmetry gap closes at $p=0.5$ in the nontrivial phase, while remaining open in the trivial phase, in agreement with Figs.~\ref{fig:dHaldane}(a,b).

\section{Conclusion}
We identify hidden critical behavior at decoherence-driven topological transitions, precisely where the topological response changes. The transition is marked by the closing of the symmetry gap, nondecaying long-range correlations, and a singular partition function in the symmetry-twisted ensemble, while the untwisted one remains gapped and conventional correlations stay short-ranged. This intrinsically mixed-state critical behavior differs fundamentally from SWSSB, as it requires only weak symmetry and can be diagnosed using single-copy observables, without the nonlinear density-matrix probes required for SWSSB. Crucially, these signatures are accessible on current quantum simulators through site- or qubit-resolved projective-measurement snapshots, providing an operational characterization of the mixed-state topological transition.

\begin{acknowledgments} 
We thank Bernhard Jobst for helpful discussions. Z.-M.~H. and S.~D. are supported by the  Deutsche Forschungsgemeinschaft (DFG, German Research Foundation) under Germany's Excellence Strategy Cluster of Excellence Matter and Light for Quantum Computing (ML4Q) EXC 2004/1 390534769 and by the DFG Collaborative Research Center (CRC) 183 Project No. 277101999 - project B02. F.~P. acknowledges support from the DFG under Germany's Excellence Strategy EXC-2111-390814868, TRR 360 (project ID 492547816), and the Munich Quantum Valley, which is supported by the Bavarian state government with funds from the Hightech Agenda Bayern Plus.
\end{acknowledgments}

\appendix 
\section{Derivation of the push-through identity Eq.~\eqref{eq:injectivity_pure}\label{app:fundamental_thm}}
For completeness, we provide details of a simplified derivation of Eq.~\eqref{eq:injectivity_pure}, reproducing the well-known fundamental theorem of matrix product states \cite{perez_garcia2007,cirac2021rmp}.

\textbf{Lemma} (Fundamental theorem for normal MPS): Let $|\psi^{\left(N\right)}\rangle$ be a normalized MPS built from a normal tensor $A$, i.e., 
\begin{equation}
|\psi^{\left(N\right)}\rangle =
	\begin{tikzpicture}[scale=.4, baseline={([yshift=0.5ex]current bounding box.center)}, thick]
		\ATensorMPS{0, 1}{$A$}
        \ATensorMPS{2, 1}{$A$}
        \draw[shift={(0,0)},dotted] (2.8,1) -- (5,1);
        \ATensorMPS{6, 1}{$A$}
        \draw[shift={(0,2.8)}] (-1.2,-1.8) arc(90:270:0.2);
         \draw[shift={(0,0)}] (7.2,0.6) arc(-90:90:0.2);
		\draw [decorate,
    	decoration = {calligraphic brace,mirror}] (-0.5,-0) --  (6.8,-0);
		\draw (2.7,-0.5) node {\scriptsize $N$};
           
	\end{tikzpicture}\ ,\ \text{with}\ N\gg 1. 
\end{equation} 
Let $|\tilde{\psi}^{(N)}\rangle$ be the same state up to a phase, represented by a different tensor $\begin{tikzpicture}[scale=.4, baseline={([yshift=-1.5ex]current bounding box.center)}, thick]
		\ATensorMPS{0, 1}{$B$}
	\end{tikzpicture}$, i.e., 
\begin{equation}
\langle \psi^{\left(N\right)}|\tilde{\psi}^{(N)}\rangle\overset{N\gg 1}{=} e^{iN\phi},
\end{equation}
then there is 
\begin{equation}
\begin{tikzpicture}[scale=.4, baseline={([yshift=-1.3ex]current bounding box.center)}, thick]
    \ATensorMPS{0,0}{$A$}
  \end{tikzpicture}=e^{-i\phi}
\begin{tikzpicture}[scale=.4, baseline={([yshift=-1.3ex]current bounding box.center)}, thick]
    \ATensorMPS{0,0}{$B$}
     \xTensor{1.6, 0}{$X$}
     \xTensor{-1.6, 0}{$Y$}
  \end{tikzpicture},\label{supp_eq:injectiveMPS_lemma}
\end{equation}
with $YX=\mathbb{1}$. 

\textbf{Proof:}
For the readers' convenience, we recall the definition of a normal MPS tensor \cite{cirac2021rmp}: A tensor $\begin{tikzpicture}[scale=.4, baseline={([yshift=-1.4ex]current bounding box.center)}, thick]
    \ATensorMPS{0,0}{$A$}
  \end{tikzpicture}$ in MPSs is called normal if there exists a finite blocking length $L_0$ such that the tensor obtained by blocking $L_0$ consecutive sites defines an injective linear map from the virtual space to the physical space, i.e.,
\begin{equation}
\begin{tikzpicture}[scale=.4, baseline={([yshift=-1.4ex]current bounding box.center)}, thick]
    \ATensorMPS{0,0}{$A^{\scalebox{0.5}{(b)}}$}
  \end{tikzpicture}\equiv
\begin{tikzpicture}[scale=.4, baseline={([yshift=0.5ex]current bounding box.center)}, thick]
		\ATensorMPS{0, 1}{$A$}
        \ATensorMPS{2, 1}{$A$}
        \draw[shift={(0,0)},dotted] (2.8,1) -- (5,1);
        \ATensorMPS{6, 1}{$A$}
       	\draw [decorate,
    	decoration = {calligraphic brace,mirror}] (-0.5,-0) --  (6.8,-0);
		\draw (2.7,-0.5) node {\scriptsize $L_0$}; 
	\end{tikzpicture}.
\end{equation}

Then, this result follows from: (i) The tensor $\begin{tikzpicture}[scale=0.4, baseline={([yshift=-1ex]current bounding box.center)}, thick]
    \ATensorMPS{0,0}{$A$}
  \end{tikzpicture}$ is normal;
  (ii) $|\tilde{\psi}^{(N)}\rangle\overset{N\gg 1}{=}e^{iN\phi}|\psi^{\left(N\right)}\rangle$. 
  
  Namely, (i) 
  implies the following decomposition of the $\begin{tikzpicture}[scale=.4, baseline={([yshift=-1.5ex]current bounding box.center)}, thick]
    \ATensorMPS{0,1}{$A$} 
  \end{tikzpicture}$ transfer matrix \cite{cirac2021rmp}, 
\begin{equation}
\begin{array}{c}
  \begin{tikzpicture}[scale=.4, baseline={([yshift=-2.4ex]current bounding box.center)}, thick]
    \ATensorMPS{0,-1}{$A$} 
    \AdagTensorMPS{0, 1}{$A^*$} ;
  \end{tikzpicture}\end{array}=
	\begin{array}{c}
		\begin{tikzpicture}[scale=.4, baseline={([yshift=-0.6ex]current bounding box.center)}, thick]
		\draw[rounded corners] (0,2.4)--(0.5,2.4)--(0.5,0.4)--(0,0.4);
        
		\draw[rounded corners] (2.2,2.4)--(1.7,2.4)--(1.7,0.4)--(2.2,0.4);

        \filldraw[fill=vector] (0.05,1.0) rectangle (0.95,1.8);
\draw (0.5,1.4) node {\scriptsize $\sigma_{\scalebox{0.5}{R}}$};
		\end{tikzpicture}
	\end{array} +\delta T(A),\ \text{with}\ 
    	\begin{array}{c}
		\begin{tikzpicture}[scale=.4, baseline={([yshift=-0.6ex]current bounding box.center)}, thick]
		\draw[rounded corners] (1,2.4)--(0.5,2.4)--(0.5,0.4)--(1,0.4);
		
		\draw[rounded corners] (1.2,2.4)--(1.7,2.4)--(1.7,0.4)--(1.2,0.4);

        \draw (1, 2.4)--(1.2, 2.4);
        \draw (1, 0.4)--(1.2, 0.4);

        \filldraw[fill=vector] (1.25,1.0) rectangle (2.15,1.8);
\draw (1.7,1.4) node {\scriptsize $\sigma_{\scalebox{0.5}{R}}$};
		\end{tikzpicture}=1
        \end{array},
    \label{supp_eq:AAbar}
\end{equation}
 where $\delta T(A)$ has a spectral radius smaller than $1$, while $\sigma_R$ is full-rank, hermitian and thus invertible. 
 
Similarly, (ii) requires 
\begin{equation}
\begin{array}{c}
  \begin{tikzpicture}[scale=.4, baseline={([yshift=-2.4ex]current bounding box.center)}, thick]
    \ATensorMPS{0,-1}{$B$} 
    \AdagTensorMPS{0, 1}{$A^*$} ;
  \end{tikzpicture}\end{array}=e^{i\phi}
	\begin{array}{c}
		\begin{tikzpicture}[scale=.4, baseline={([yshift=-0.6ex]current bounding box.center)}, thick]
		\draw[rounded corners] (0,2.4)--(0.5,2.4)--(0.5,0.4)--(0,0.4);
        
		\draw[rounded corners] (2.2,2.4)--(1.7,2.4)--(1.7,0.4)--(2.2,0.4);
        

        \filldraw[fill=vector] (-.05,1.0) rectangle (0.85,1.8);
        \draw (0.4,1.4) node {\scriptsize $X_{\scalebox{0.5}{R}}$};
        
        \filldraw[fill=vector] (1.25,1.0) rectangle (2.15,1.8);
        \draw (1.7,1.4) node {\scriptsize $X_{\scalebox{0.5}{L}}$};
        
		\end{tikzpicture}
	\end{array} + \delta T(B),
    \ \text{with}\ 
    	\begin{array}{c}
		\begin{tikzpicture}[scale=.4, baseline={([yshift=-0.6ex]current bounding box.center)}, thick]
		\draw[rounded corners] (1,2.4)--(0.5,2.4)--(0.5,0.4)--(1,0.4);
		
		\draw[rounded corners] (1.2,2.4)--(1.7,2.4)--(1.7,0.4)--(1.2,0.4);

        \filldraw[fill=vector] (-.05,1.0) rectangle (0.85,1.8);
        \draw (0.4,1.4) node {\scriptsize $X_{\scalebox{0.5}{$L$}}$};

           \filldraw[fill=vector] (1.15,1.0) rectangle (2.05,1.8);
        \draw (1.6,1.4) node {\scriptsize $X_{\scalebox{0.5}{$R$}}$};

        \draw (1, 2.4)--(1.2, 2.4);
        \draw (1, 0.4)--(1.2, 0.4);
		\end{tikzpicture}=1
        \end{array},   
\end{equation}
 where $\delta T(B)$ has a spectral radius smaller than $1$, and  $X_{R, L}$ might not be invertible. One can derive this spectral property directly via the Jordan normal form.
 
Built upon these, we obtain the following two key identities, \begin{subequations}\label{supp_eq:2id} 
  \begin{equation}\label{supp_eq:2id:a} 
    \begin{array}{c}
      \begin{tikzpicture}[scale=.4, baseline={([yshift=-3.2ex]current bounding box.center)}, thick]
        \ATensorMPS{(0, 0)}{$A$}
        \xTensor{(1.6, 0)}{$\sigma_{\scalebox{0.5}{$R$}}^{\scalebox{0.5}{$T$}}$}
      \end{tikzpicture}
    \end{array}
    = e^{-i\phi}
    \begin{tikzpicture}[scale=.4, baseline={([yshift=-1.3ex]current bounding box.center)}, thick]
      \ATensorMPS{(0, 0)}{$B$}
      \xTensor{(1.6, 0)}{$X_{\scalebox{0.5}{$R$}}^{\scalebox{0.5}{$T$}}$}
      \xTensor{(-1.6, 0)}{$X_{\scalebox{0.5}{$L$}}$}
    \end{tikzpicture},
  \end{equation}
and
  \begin{equation}\label{supp_eq:2id:b} 
    \begin{array}{c}
      \begin{tikzpicture}[scale=.4, baseline={([yshift=-0.6ex]current bounding box.center)}, thick]
        \xTensor{(1.6, 0)}{$\sigma_{\scalebox{0.5}{$R$}}^{\scalebox{0.5}{$T$}}$}
      \end{tikzpicture}
    \end{array}
    =
    \begin{tikzpicture}[scale=.4, baseline={([yshift=-0.6ex]current bounding box.center)}, thick]
      \xTensor{(-0.8, 0)}{$X_{\scalebox{0.5}{$L$}}$}
      \xTensor{(0.8, 0)}{$X_{\scalebox{0.5}{$R$}}^{\scalebox{0.5}{$T$}}$}
    \end{tikzpicture},
  \end{equation}
\end{subequations}
by contracting $|\psi^{\left(N\right)} \rangle$ and $|\tilde{\psi}^{(N)} \rangle$ with $A^*$ over $N-1$ and $N$ sites ($N\gg 1$), respectively, i.e., 
\begin{subequations}\label{supp_eq:N-1_contract}
\begin{align}
\text{(a)}\quad
\begin{array}{c}
\begin{tikzpicture}[scale=.4, baseline={([yshift=0.6ex]current bounding box.center)}, thick]
  \draw[shift={(0,0)},dotted] (1,1) -- (5,1);
  \ATensorMPS{0, 1}{$A$}
  \ATensorMPS{4.2, 1}{$A$}
  \ATensorMPS{6, 1}{$A$}
  \draw[shift={(0,2.8)}] (-1.2,-1.8) arc(90:270:0.2);
  \draw[shift={(0,0)}] (7.2,0.6) arc(-90:90:0.2);
  \draw [decorate,decoration = {calligraphic brace,mirror}] (-0.5,-0) -- (6.8,-0);
  \draw (2.7,-0.5) node {\scriptsize $N$};
  \draw[shift={(0,0)},dotted] (1,2.8) -- (5,2.8);
  \AdagTensorMPS{0, 2.8}{$A^*$}
  \AdagTensorMPS{4.2, 2.8}{$A^*$}
\end{tikzpicture}
\end{array}
&\overset{N\rightarrow \infty}{=}
\begin{array}{c}
\begin{tikzpicture}[scale=.4, baseline={([yshift=-1.2ex]current bounding box.center)}, thick]
  \ATensorMPS{(0, 0)}{$A$}
  \xTensor{(1.6, 0)}{$\sigma_{\scalebox{0.5}{$R$}}^{\scalebox{0.5}{$T$}}$}
\end{tikzpicture}
\end{array},
\label{supp_eq:N-1_a}
\\[6pt]
\text{(b)}\quad
\begin{array}{c}
\begin{tikzpicture}[scale=.4, baseline={([yshift=0.6ex]current bounding box.center)}, thick]
  \draw[shift={(0,0)},dotted] (1,1) -- (5,1);
  \ATensorMPS{0, 1}{$B$}
  \ATensorMPS{4.2, 1}{$B$}
  \ATensorMPS{6, 1}{$B$}
  \draw[shift={(0,2.8)}] (-1.2,-1.8) arc(90:270:0.2);
  \draw[shift={(0,0)}] (7.2,0.6) arc(-90:90:0.2);
  \draw [decorate,decoration = {calligraphic brace,mirror}] (-0.5,-0) -- (6.8,-0);
  \draw (2.7,-0.5) node {\scriptsize $N$};
  \draw[shift={(0,0)},dotted] (1,2.8) -- (5,2.8);
  \AdagTensorMPS{0, 2.8}{$A^*$}
  \AdagTensorMPS{4.2, 2.8}{$A^*$}
\end{tikzpicture}
\end{array}
&\overset{N\rightarrow \infty}{=}
\scalebox{0.7}{$e^{i(N-1)\phi}$}\;
\begin{array}{c}
\begin{tikzpicture}[scale=.4, baseline={([yshift=-1.3ex]current bounding box.center)}, thick]
  \ATensorMPS{(0, 0)}{$B$}
  \xTensor{(1.6, 0)}{$X_{\scalebox{0.5}{$R$}}^{\scalebox{0.5}{$T$}}$}
  \xTensor{(-1.6, 0)}{$X_{\scalebox{0.5}{$L$}}$}
\end{tikzpicture}
\end{array},
\label{supp_eq:N-1_b}
\end{align}
\end{subequations}

and 
\begin{subequations}\label{supp_eq:N_contract}
\begin{align}
\text{(a)}\quad
\begin{array}{c}
\begin{tikzpicture}[scale=.4, baseline={([yshift=0.6ex]current bounding box.center)}, thick]
  \draw[shift={(0,0)},dotted] (1,1) -- (5,1);
  \ATensorMPS{0, 1}{$A$}
  \ATensorMPS{4.2, 1}{$A$}
  \ATensorMPS{6, 1}{$A$}
  \draw[shift={(0,2.8)}] (-1.2,-1.8) arc(90:270:0.2);
  \draw[shift={(0,0)}] (7.2,0.6) arc(-90:90:0.2);
  \draw [decorate, decoration = {calligraphic brace,mirror}] (-0.5,-0) -- (6.8,-0);
  \draw (2.7,-0.5) node {\scriptsize $N$};
  \draw[shift={(0,0)},dotted] (1,2.8) -- (5,2.8);
  \AdagTensorMPS{0, 2.8}{$A^*$}
  \AdagTensorMPS{4.2, 2.8}{$A^*$}
  \AdagTensorMPS{6, 2.8}{$A^*$}
\end{tikzpicture}
\end{array}
&\overset{N\rightarrow \infty}{=}
\begin{array}{c}
\begin{tikzpicture}[scale=.4, baseline={([yshift=-0.4ex]current bounding box.center)}, thick]
  \xTensor{0,0}{$\sigma_{\scalebox{0.5}{$R$}}^{\scalebox{0.5}{$T$}}$}
\end{tikzpicture}
\end{array},
\label{supp_eq:N_a}
\\[6pt]
\text{(b)}\quad
\begin{array}{c}
\begin{tikzpicture}[scale=.4, baseline={([yshift=0.6ex]current bounding box.center)}, thick]
  \draw[shift={(0,0)},dotted] (1,1) -- (5,1);
  \ATensorMPS{0, 1}{$B$}
  \ATensorMPS{4.2, 1}{$B$}
  \ATensorMPS{6, 1}{$B$}
  \draw[shift={(0,2.8)}] (-1.2,-1.8) arc(90:270:0.2);
  \draw[shift={(0,0)}] (7.2,0.6) arc(-90:90:0.2);
  \draw [decorate, decoration = {calligraphic brace,mirror}] (-0.5,-0) -- (6.8,-0);
  \draw (2.7,-0.5) node {\scriptsize $N$};
  \draw[shift={(0,0)},dotted] (1,2.8) -- (5,2.8);
  \AdagTensorMPS{0, 2.8}{$A^*$}
  \AdagTensorMPS{4.2, 2.8}{$A^*$}
  \AdagTensorMPS{6, 2.8}{$A^*$}
\end{tikzpicture}
\end{array}
&\overset{N\rightarrow \infty}{=}
\scalebox{0.8}{$e^{iN\phi}$}\;
\begin{array}{c}
\begin{tikzpicture}[scale=.4, baseline={([yshift=-0.4ex]current bounding box.center)}, thick]
  \xTensor{(0.8, 0)}{$X_{\scalebox{0.5}{$R$}}^{\scalebox{0.5}{$T$}}$}
  \xTensor{(-0.8, 0)}{$X_{\scalebox{0.5}{$L$}}$}
\end{tikzpicture}
\end{array}.
\label{supp_eq:N_b}
\end{align}
\end{subequations}

Finally, from these two identities (Eq.~\eqref{supp_eq:2id}), we reproduce Eq.~\eqref{supp_eq:injectiveMPS_lemma}, i.e., 
\begin{equation}
\begin{cases}
Y\equiv X_{L}\\
X\equiv X_{R}^{T}\ \frac{1}{\sigma_{R}^{T}}
\end{cases}\ \overset{\eqref{supp_eq:2id:b}}{\implies} Y X =\mathbb{1}.
\end{equation}

\subsection{Application to onsite unitary transformations}
We now apply this result to states related by an onsite unitary transformation. We can show that $X$ is invertible, so that $Y=X^{-1}$. Additionally, $X$ can be chosen to be unitary, e.g., by choosing the canonical gauge in which the transfer matrix is unital~\cite{cirac2021rmp}.
\\

\textbf{Observation 1: }Let $|\psi^{\left(N\right)}\rangle$ be an MPS built from a normal tensor $A$, i.e., 
\begin{equation}
|\psi^{\left(N\right)}\rangle =
	\begin{tikzpicture}[scale=.4, baseline={([yshift=0.5ex]current bounding box.center)}, thick]
		\ATensorMPS{0, 1}{$A$}
        \ATensorMPS{2, 1}{$A$}
        \draw[shift={(0,0)},dotted] (2.8,1) -- (5,1);
        \ATensorMPS{6, 1}{$A$}
        \draw[shift={(0,2.8)}] (-1.2,-1.8) arc(90:270:0.2);
         \draw[shift={(0,0)}] (7.2,0.6) arc(-90:90:0.2);
		\draw [decorate,
    	decoration = {calligraphic brace,mirror}] (-0.5,-0) --  (6.8,-0);
		\draw (2.7,-0.5) node {\scriptsize $N$};
           
	\end{tikzpicture}\ ,\ \text{with}\ N\gg 1, 
\end{equation}
and $U=u^{\otimes N}$ an on-site operator, such that, 
\begin{equation}
u^{\otimes N}|\psi^{\left(N\right)}\rangle=e^{i N \phi}|\psi^{\left(N\right)}\rangle.
\label{supp_eq:equality_mps}
\end{equation}
Then there is
\begin{equation}
\begin{tikzpicture}[scale=.4, baseline={([yshift=-1.3ex]current bounding box.center)}, thick]
    \ATensorMPS{0,0}{$A$}
  \end{tikzpicture}=e^{-i\phi}
\begin{tikzpicture}[scale=.4, baseline={([yshift=-2.6ex]current bounding box.center)}, thick]
    \ATensorMPS{0,0}{$A$}
    \wTensor{0,0.9}{$u$}
     \xTensor{1.6, 0}{$X$}
     \xTensor{-1.6, 0}{$X^{\scalebox{0.5}{-1}}$}
  \end{tikzpicture},\ \text{with}\ X\ \text{invertible}.\label{supp_eq:MPS_injective_invertible}
\end{equation}

\textbf{Proof:} This follows directly from the Lemma above, where $X$ and $Y$ are both square matrices, i.e.,
\begin{equation}
YX=\mathbb{1}\implies \det Y\det X\neq 0,
\end{equation}
so $Y$ is invertible and thus $Y=X^{-1}$
reproducing Eq.~\eqref{supp_eq:MPS_injective_invertible}.

\begin{figure}
\includegraphics[scale=0.035]{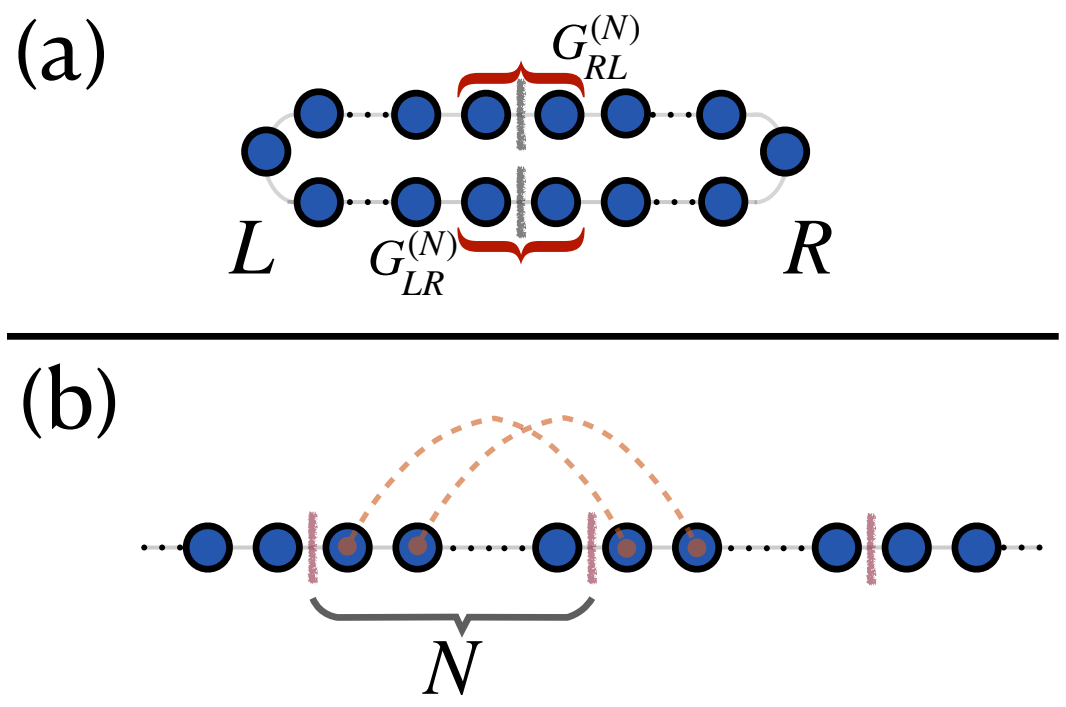}\caption{Illustration of flux insertion on a finite periodic chain. (a) Local terms in the modular Hamiltonian $G^{(N)}$. (b) The corresponding $\mathbb{Z}/N\mathbb{Z}$ quotient construction.
\label{appfig:flux}}
\end{figure}
        
\section{Review of flux insertion \label{app_sec:flux_insertion}}
In this section, we illustrate the insertion of a symmetry flux into the density matrix and its tensor-network representation, following Refs.~\cite{huang2025prl_mspt,huang2026arxiv}.

We start by reviewing the usual implementation of flux insertion for a periodic $N$-site chain, in the density matrix representation. Consider $G^{\left(N\right)}\equiv-\ln\rho^{\left(N\right)}$ a local modular Hamiltonian with the superscript $N$ emphasizing the finite system size. Locality is assumed only for convenience, as it allows the flux to be inserted directly at the level of the modular Hamiltonian, while the more general construction developed below does not rely on this assumption.
To this end, we decompose $G^{(N)}$ by choosing a cut, rendering [see Fig. \ref{appfig:flux}(a)]
\begin{equation}
G^{\left(N\right)}=G^{\left(N\right)}_{L}+G^{\left(N\right)}_{R}+{G}^{\left(N\right)}_{LR}+G^{\left(N\right)}_{RL},
\end{equation}
where the subscript $L$ and $R$ denote terms supported within the left and right intervals, while $LR$ and $RL$ for those cross the corresponding boundaries. The flux-inserted $G^{(N)}$ is then given as 
\begin{equation}
G^{\left(N\right)}\left(\boldsymbol{\phi}_{g}\right)=G^{\left(N\right)}_{L}+G^{\left(N\right)}_{R}+U_{g_1,\ R}G^{\left(N\right)}_{LR}U^{\dagger}_{g_1,\ R}+G^{\left(N\right)}_{RL},\label{app_eq:modular_Ham_finite}
\end{equation}
where $U_{g_1,R}$ acts only on the right-supported degrees of freedom of $G^{\left(N\right)}_{LR}$, rendering a twisted boundary term.

In the tensor-network representation, this construction applies to a general density matrix, independent of its microscopic form. One views the finite periodic chain as an $N$-site unit cell of a translationally invariant infinite chain and identifies its endpoints by taking the quotient $\mathbb{Z}/(N\mathbb{Z})$~\cite{zaletel2013prl,zaletel2014njp} [see Fig.~\ref{appfig:flux}(b)]. On the infinite chain, the twisted boundary term is implemented by applying $u_{g_1}^n$ to every site in the segment $[nN+1,(n+1)N]$, for each integer $n$. Restricting this transformed infinite system to a single $N$-site unit cell then yields the flux-inserted density matrix,
\begin{equation}
\rho(V_{g_1})=
\begin{array}{c}
	\begin{tikzpicture}[scale=.45, baseline={([yshift=0ex]current bounding box.center)}, thick]
		\draw[shift={(0,0)},dotted] (0,0) -- (6,0);
        \draw[shift={(0,0)},dotted] (0,2) -- (6,2);
        \draw[shift={(0,0)}] (-2.9,0) arc(90:270:0.2);
        \draw[shift={(0,2)}] (-2.9,0) arc(90:270:0.2);
        \draw[shift={(0,0)}] (6.6,-0.4) arc(-90:90:0.2);
        \draw[shift={(0,2)}] (6.6,-0.4) arc(-90:90:0.2);
		\ATensorLPDO{0.4,2}{$A$}
        \AdagTensorLPDO{0.4,0}{$A^*$}
        \xTensor{-1.9,0}{$V_{g_1}^*$}
		\ATensorLPDO{5,2}{$A$}
        \AdagTensorLPDO{5,0}{$A^*$}
        \xTensor{-1.9,2}{$V_{g_1}$}
        \draw[red] (1.7,-1.2)--(1.7,3.2);
        \draw[red] (6.3,-1.2)--(6.3,3.2);
        \draw[red] (1.7,3.2) arc(0:180:0.35);
        \draw[red] (6.3,3.2) arc(0:180:0.35);
        \draw[red] (1.7,-1.2) arc(0:-180:0.35);
        \draw[red] (6.3,-1.2) arc(0:-180:0.35);
        \draw[preaction={draw=white, line width=5pt}] (1.5,0) --(2,0);
        \draw[preaction={draw=white, line width=5pt}] (1.5,2) --(2,2);
        \draw[preaction={draw=white, line width=5pt}] (6.1,0) --(6.6,0);
        \draw[preaction={draw=white, line width=5pt}] (6.1,2) --(6.6,2);

	\end{tikzpicture}
	\end{array}.
\end{equation}
\\

\bibliographystyle{apsrev4-2}
\bibliography{AMWF}

\end{document}